\documentclass[a4paper,11pt]{article}
\pdfoutput=1 
\usepackage{jheppub} 
\usepackage{array}
\usepackage{multirow}
\usepackage{amsmath}
\usepackage{slashed}
\usepackage{amstext,amssymb}
\usepackage{graphicx}
\usepackage{graphicx,wrapfig,lipsum}
\usepackage{xmpmulti}
\usepackage{animate}
\usepackage{epstopdf}
\usepackage{xcolor,colortbl}
\usepackage{multimedia}
\usepackage{hyperref}
\usepackage{url}
\usepackage{subcaption}
\usepackage{dsfont}
\usepackage{xspace}
\usepackage{color}
\usepackage{units}
\usepackage{braket}
\usepackage{makecell}
\usepackage{bigstrut}
\usepackage{cleveref}
\usepackage{caption}
\usepackage{tikz}
\usepackage{wrapfig}
\usetikzlibrary{snakes}

\title{Leptogenesis in a Left-Right Symmetric Model with double seesaw}
\author[a]{Utkarsh Patel}
\author[a]{, Pratik Adarsh}
\author[a]{, Sudhanwa Patra}
\author[a,b,c]{, Purushottam Sahu}
\affiliation[a]{Department of Physics, Indian Institute of Technology Bhilai, Kutelabhatha 491001, India}
\affiliation[b]{International Centre for Theoretical Physics (ICTP),Strada Costiera 11, Trieste 34151, Italy}
\affiliation[c]{Department of Physics, Indian Institute of Technology Bombay, India}
\emailAdd{utkarshp@iitbhilai.ac.in}
\emailAdd{pratikad@iitbhilai.ac.in}
\emailAdd{sudhanwa@iitbhilai.ac.in}
\emailAdd{purushottams@iitbhilai.ac.in}

\abstract{We explore the connection between low-scale CP-violating Dirac phase~$(\delta)$ and high-scale leptogenesis in a Left-Right Symmetric Model (LRSM) with scalar bidoublet and doublets. The fermion sector of the model is extended with one sterile neutrino~$(S_L)$ per generation to implement a double seesaw mechanism in the neutral fermion mass matrix. The double seesaw is performed via the implementation of type-I seesaw twice. The first seesaw facilitates the generation of Majorana mass term for heavy right-handed (RH) neutrinos~$(N_R)$, and the light neutrino mass becomes linearly dependent on $S_L$ mass in the second. In our framework, we have taken charge conjugation ($C$) as the discrete left-right (LR) symmetry. This choice assists in deriving the Dirac neutrino mass matrix ($M_D$) in terms of the light and heavy RH neutrino masses and light neutrino mixing matrix $U_{PMNS}$ (containing $\delta$). We illustrate the viability of unflavored thermal leptogenesis via the decay of RH neutrinos by using the obtained $M_D$ with the masses of RH neutrinos as input parameters. A complete analysis of the Boltzmann equations describing the asymmetry evolution is performed in the unflavored regime, and it is shown that with or without Majorana phases, the CP-violating Dirac phase is sufficient to produce the required asymmetry in the leptonic sector within this framework for a given choice of input parameters. Finally, we comment on the possibility of constraining our model with the current and near-future oscillation experiments, which are aimed at refining the value of $\delta$.}


\begin{document} 
	\maketitle
	\flushbottom
	\section{Introduction}
	\label{sec:intro}
	LRSM \cite{Mohapatra:1974gc,Pati:1974yy} was conceived as a theoretical extension of the Standard Model (SM) of particle physics where the left- and right-handed fermion fields are treated on a similar footing. It, therefore, naturally explains the observed parity violations in weak interactions through the dynamics of symmetry breaking \cite{Senjanovic:1975rk,Senjanovic:1978ev}. The results from neutrino oscillation experiments \cite{SNO:2002tuh,Super-Kamiokande:2016yck,T2K:2019efw,DayaBay:2012fng,DoubleChooz:2011ymz} necessitate the existence of non-zero masses for neutrinos. However, the SM predicts massless neutrinos and this problem is generally tackled by extending the SM to achieve seesaw mechanism. This mechanism has various types depending on the particle considered in the SM extension. Among them, we have type-I \cite{Minkowski:1977sc, Mohapatra:1979ia, Yanagida:1979as, GellMann:1980vs}, type-II \cite{Magg:1980ut, Schechter:1980gr, Cheng:1980qt, Lazarides:1980nt, Mohapatra:1980yp} and type-III \cite{Foot:1988aq,Ma:2001kg,Ma:2002pf,Barr:2005je,Dorsner:2006fx,He:2012ub} seesaws, which can be realized by extending the SM with a right-handed neutrino, a scalar triplet and a fermion triplet, respectively. The LRSM encompasses the full spectrum of chiral fermions, which includes right-handed neutrinos ($N_R$) that can have both Dirac and Majorana mass terms depending on the scalar content of the model. The choice of scalars also regulates the phenomenological aspects of the model. For LRSM models featuring Higgs doublets, neutrinos acquire solely Dirac masses. Conversely, in LRSM models involving Higgs triplets, Majorana masses are generated, leading to light neutrino masses through a combined type-I and type-II seesaw mechanism. The later version of LRSM is generally more favoured as it facilitates lepton number violation (LNV) through the Majorana mass terms which can lead to leptogenesis \cite{Fukugita:1986hr}.
	
	Leptogenesis has been a widely studied phenomenon in particle physics that attempts to account for the observed asymmetry in the matter-antimatter content of the Universe, which has been an essential missing feature of the SM. In leptogenesis, an asymmetry in the lepton/antilepton number density is created before the electroweak phase transition, which via the higher dimensional anomalous $B+L$ violating sphelaron processes~\cite{Klinkhamer:1984di,Arnold:1987mh,Kuzmin:1985mm,Rubakov:1996vz} is converted into the observed baryon asymmetry of the Universe (BAU). One of the early realizations of leptogenesis, as proposed in \cite{Fukugita:1986hr}, was the CP-violating and lepton number violating out-of-equilibrium decay of heavy neutrinos introduced as singlets under the SM gauge group. Such a type of leptogenesis is referred to as thermal leptogenesis in the literature, and here the typical masses of right-handed neutrinos are taken to be very high at around $10^{10}$ GeV or more. Thermal leptogenesis can further be classified as unflavored or flavored, depending on whether we consider flavor mixings within the leptonic sectors. As leptogenesis is an early Universe phenomenon, it depends on high-energy parameters which could be inaccessible to probe in current or future experiments. Thus, several attempts have been made to connect leptogenesis and low-energy CP violation directly. For example, in Ref.~\cite{Branco:2001pq}, the relation between CP-violating phases at high- and low-energy is shown to exist in the framework of minimal seesaw embedded in a specific grand unified model with weak-basis (structured to have all gauge currents as real and flavour diagonal). References \cite{Molinaro:2009lud,Xing:2020erm,Granelli:2021fyc} adopt Casas-Ibarra (CI) parametrization \cite{Casas:2001sr} to relate the Dirac neutrino mass matrix or neutrino Yukawa coupling matrix with light neutrino mixing matrix (containing low-energy CP-violating phases). In Ref. \cite{Molinaro:2009lud}, the low- and high-energy CP violation is shown to be connected with an orthogonal matrix (from CI parametrization) with unknown phases. The paper investigates the interplay between low and high-energy CP-violating phases originating from the light neutrino mixing matrix and the orthogonal matrix to study flavored leptogenesis. Ref. \cite{Xing:2020erm} realizes a direct connection between unflavored thermal leptogenesis and CP violation at low-energies based on the quantum corrected CI parametrization of $M_D$ with the help of one-loop renormalization group equations (RGEs) between the seesaw and electroweak (EW) scales. Ref. \cite{Granelli:2021fyc} studies the transition between the different flavor regimes in high-scale leptogenesis based on type-I seesaw mechanism with low-energy CP-violating phases. The orthogonal matrix from CI parametrization of $M_D$ is shown to depend only on one complex angle and can be considered purely real or imaginary for different light neutrino mass spectrum hierarchies. Thus, the leptogenesis with low-energy CP-violating phases without CP violation associated with the orthogonal matrix is achieved. References \cite{Joshipura:2001ya,Rodejohann:2002mh,Babu:2005bh,Akhmedov:2006yp,Chao:2007rm,Hallgren:2007nq,Abada:2008gs,Rink:2020uvt,Granelli:2023tcj} are some other notable works along the similar lines. Analogous to CI parametrization (orthogonal parametrization), there are other parametrizations for Dirac neutrino mass matrix to correlate high- and low-energy CP-violating phases \cite{Pascoli:2003uh,Branco:2011zb,Joshipura:2001ui,Davidson:2001zk,Rahat:2020mio}.
	
	In the context of LRSM with doublet scalars, we see that originally, there were no Majorana mass terms possible for heavy right-handed neutrinos, thus prohibiting leptogenesis. An extension of the fermion sector with three generations of sterile neutrino, $S_L$ facilitates us to perform a double type-I seesaw \cite{PhysRevLett.56.561, PhysRevD.34.1642} in neutrino mass matrix to provide a Majorana nature to right-handed neutrinos. This framework has been implemented for studying neutrinoless double beta decay in Ref. \cite{Patra:2023ltl}. We work in a similar framework to study thermal unflavored leptogenesis and to establish a direct link between low- and high-energy CP violations. In Ref. \cite{Zhang:2020lir}, similar work has been done in the Minimal Left-Right Symmetric Model (MLRSM) with scalar triplets. Our work differs in the context that the model framework and the approach to derive the structure of the Dirac mass matrix ($M_D$) are unlike them. The difference is that instead of employing the CI parametrization to find the structure of $M_D$, the theoretical framework with considered particle spectrum incorporated with discrete LR symmetry $C$ and the screening condition (the condition for cancellation of Dirac structures in the expression of light neutrino mass matrix) naturally gives us $M_D$ in terms of low-energy oscillation parameters.
	
	The key point that we focus in our numerical analysis is the dependence of BAU on the low-scale CP-violating phase $(\delta)$. The derived coincidence of light and heavy neutrino mixings in our framework allows the dynamical evolution of baryon sector asymmetry (due to the decay of heavy right-handed neutrinos) to depend on the light neutrino sector. Such a connection not only reduces the number of input parameters but also strongly enhances the viability of our model.
	
	The paper is structured in the following manner. In section~\ref{sec:framework_motivation}, we introduce the model framework for our work and also list the particle content and their basic interactions. The double seesaw structure of the neutrino mass matrix is also presented in a subsection. In the subsequent section~\ref{sec:connec_High_Low}, we discuss the connection between low- and high-scale physics, and also derive the structures of Dirac mass matrix, $M_D$ and asymmetry parameter~$(\epsilon_1)$. Next, we present our benchmark analysis for leptogenesis in section~\ref{sec:lepto} by sketching the full Boltzmann numerical solutions of asymmetry evolution. Interdependencies of $\epsilon_1$ and CP-phase $(\delta)$ on each-other and on other model parameters are also explored here. Finally, in section~\ref{sec:conc} we conclude our work by summarizing the key results from our analysis and also comment on the scope of future work. All the other relevant details and calculations can be found in the appendix.
	
	\section{Model framework and motivation}\label{sec:framework_motivation}
	The SM explains the $(V-A)$ structure of the weak interaction and parity violation through the SM gauge group $\mathcal{G}\equiv SU(3)_C\times SU(2)_L \times U(1)_Y$. This is evident as all right-handed fields transform trivially under $SU(2)_L$. However the SM does not explain the origin of the parity violation. It is, therefore, natural to seek an explanation for parity violation from a theory that conserves parity at a higher energy scale. This led to the development of an extension of the SM gauge theory known as the LRSM. In the LRSM, the SM gauge group is extended to 
	\begin{equation}
		\mathcal{G}_{LR} \equiv SU(3)_C \times SU(2)_L \times SU(2)_R \times U(1)_{B-L}
	\end{equation}
	where $B - L$ represents the difference between baryon (B) and lepton (L) numbers. The electric charge $Q$ is defined as
	\begin{equation}
		Q=T_{3L}+T_{3R}+\frac{B-L}{2}.
	\end{equation}
	Here, $T_{3L}$ and $T_{3R}$ are, respectively, the third components of isospin of the gauge groups $SU(2)_L$ and $SU(2)_R$. The fermion spectrum of this model comprises all the SM fermions plus a right-handed neutrino $N_R$. The fermion fields with their quantum numbers ($SU(2)_L \times SU(2)_R \times U(1)_{B-L}$) are as follows:
	\begin{eqnarray}
		&&q_{L}=\begin{pmatrix}u_{L}\\
			d_{L}\end{pmatrix}\equiv[2,1,1/3]\,, ~ q_{R}=\begin{pmatrix}u_{R}\\
			d_{R}\end{pmatrix}\equiv[1,2,1/3]\,,\nonumber \\
		&&\ell_{L}=\begin{pmatrix}\nu_{L}\\
			e_{L}\end{pmatrix}\equiv[2,1,-1] \,, ~\,\, \ell_{R}=\begin{pmatrix} N_{R}\\
			e_{R}\end{pmatrix}\equiv[1,2,-1] \,. \nonumber
	\end{eqnarray}
	We have dropped the $SU(3)_C$ quantum numbers for simplicity. The model's scalar sector comprises Higgs bidoublet $\Phi$ and the Higgs doublets: $H_L$ and $H_R$. The matrix structures of the scalar fields are,
\begin{equation*}
		\Phi=
	\begin{pmatrix} 
		\phi_{1}^0     &  \phi_{2}^+ \\
		\phi_{1}^-     &  \phi_{2}^0
	\end{pmatrix} \sim [2,2,0],~~ H_L=
\begin{pmatrix} 
h_L^+\\
h_L^0
\end{pmatrix}\sim[2,1,1],~~	H_R=
\begin{pmatrix} 
h_R^+  \\
h_R^0
\end{pmatrix}\sim[1,2,1].
\end{equation*}
	The spontaneous symmetry breaking (SSB) scheme from LRSM to SM to $U(1)_{em}$ is as follows:
	\begin{center}
		\textbf{\underline{SSB of LRSM}}
	\end{center}
	\begin{equation*}
		SU(2)_L \times \underbrace{SU(2)_R \times U(1)_{B-L}}
	\end{equation*}
	\begin{equation*}
		\hspace{4cm}	\Big\downarrow \braket{H_R(1,2,1)}
	\end{equation*}
	\begin{equation*}
		\hspace{0.8cm}	\underbrace{SU(2)_L \times U(1)_Y}
	\end{equation*}
	\begin{equation*}
		\hspace{5.2cm}	\Big\downarrow \braket{\phi(1_L,1/2_Y)}\subset \Phi(2,2,0)
	\end{equation*}
	\begin{equation*}
		\hspace{1.5cm}	U(1)_{em}
	\end{equation*}
	At the LRSM scale, the neutral component $h_R^0$ of Higgs doublet $H_R$ acquires a vacuum expectation value (VEV) $\braket{H_R^0}\equiv v_R$ that causes SSB from LRSM to SM. The right-handed gauge bosons $W_{R}^{\pm}$ and $Z'$ get their masses from this symmetry breaking and depend on the breaking scale ($v_R$). The Higgs doublet $H_L$ does not participate in SSB, but it is required in the particle spectrum for left-right invariance. The electroweak symmetry breaking ($SU(2)_L\times U(1)_Y \rightarrow U(1)_{em}$) is achieved by assigning non-zero VEVs: $\braket{\phi_{1}^0}\equiv v_1$ and $\braket{\phi_{2}^0}\equiv v_2$ to the neutral components of Higgs bidoublet $\Phi$, with $v=\sqrt{v_1^2+v_2^2}\simeq 246$ GeV. The Yukawa Lagrangian with usual quarks and leptons reads as,
	\begin{eqnarray}\label{Lag1}
		-\mathcal{L}_{Yuk} &\supset& \overline{q_{L}} \left[Y_1 \Phi + Y_2 \widetilde{\Phi} \right] q_R 
		+\,\overline{\ell_{L}} \left[Y_3 \Phi + Y_4 \widetilde{\Phi} \right] \ell_R+ \mbox{h.c.}\,.
		\label{YukLag}
	\end{eqnarray}
	Here $\widetilde{\Phi} = \sigma_2 \Phi^* \sigma_2$ and $\sigma_2$ is the second Pauli matrix. When the neutral components ($\phi_{1}^0$ and $\phi_{2}^0$) of Higgs bidoublet acquire non-zero VEVs,
	\begin{equation}
		\langle \Phi \rangle
		=
		\begin{pmatrix}
			v_1 & 0 \\
			0 & v_2 
		\end{pmatrix}\,,
	\end{equation}
	they give masses to quarks, charged leptons and light neutrinos (Dirac mass) as follows,
	\begin{eqnarray}
		&&  M_u =  Y_1 v_1 + Y_2 v_2\,, \quad \quad \nonumber \\
		&&    M_d =  Y_1 v_2 + Y_2 v_1\,, \quad \quad \nonumber \\
		&&  M_e =  Y_3 v_2 + Y_4 v_1 \,,\quad \quad\nonumber \\
		&& M^\nu_D\equiv M_D = Y_3 v_1 + Y_4 v_2\,.
	\end{eqnarray}
	Here, $M_u$ and $M_d$ are the up-type and down-type quark mass matrices, and $M_e$ and $M_D$ represent charged lepton and Dirac neutrino mass matrices, respectively. In contrast to Yukawa couplings, which are complex for non-zero CP-asymmetry \cite{Covi:1996wh}, VEVs $v_1$ and $v_2$ are here assumed to be real. If we have $v_2 \ll v_1$ and $|Y_3|\ll |Y_4|$, this will result in small Dirac neutrino masses. With these assumptions, the charged lepton and light neutrino masses can be re-expressed as:
	\begin{eqnarray}
			M_e &\approx & Y_4 v_1\,, \\
			M_D &=& v_1 \left( Y_3 + M_e  \frac{v_2}{v_1^2} \right)\approx v Y_3\equiv vY_D\,.\label{MDY3} 
	\end{eqnarray}
	Here, $v=\sqrt{v_1^2+v_2^2}\approx v_1$ for $v_2 \ll v_1$.The Left-Right Symmetric Model we have discussed above is minimal and based on gauge groups $ SU(2)_L \times SU(2)_R \times U(1)_{B-L}$ with coupling constants $g_L$, $g_R$ and $g_{BL}$ respectively. For the left and right gauge coupling constants to be equal, i.e. $g_L=g_R$, we need an additional discrete LR symmetry. The choice of this LR symmetry is twofold \cite{Maiezza:2010ic}: (i.) a generalized parity $\mathcal{P}$ and (ii.) a generalized charge conjugation $\mathcal{C}$. Under the parity and charge conjugation operations, the fields transform as follows:
	\begin{equation}
		\mathcal{P}:\left\{
		\begin{aligned}
			& \ell_L \leftrightarrow \ell_R, \quad q_L\leftrightarrow q_R, \\
			& \Phi \leftrightarrow \Phi^{\dagger}, \quad H_L \leftrightarrow H_R, \quad \tilde{\Phi} \leftrightarrow \tilde{\Phi}^{\dagger}\\
			&
		\end{aligned}\quad\right |
		\quad\mathcal{C}:\left\{
		\begin{aligned}
			& \ell_L \leftrightarrow \ell_R^c, \quad q_L\leftrightarrow q_R^c, \\
			& \Phi \leftrightarrow \Phi^{T}, \quad H_L \leftrightarrow H_R^*, \quad \tilde{\Phi} \leftrightarrow \tilde{\Phi}^{T}. \\
			&
		\end{aligned}\right.
	\end{equation}
	Imposition of either of these discrete symmetries in LRSM makes the Lagrangian in Eq. (\ref{Lag1}) invariant, and it leads to Hermitian Yukawa matrices for the case of discrete $\mathcal{P}$ symmetry. For the case of discrete $\mathcal{C}$ symmetry, the Yukawa matrices become symmetric. In our discussion, we consider $\mathcal{C}$ symmetry as the additional discrete symmetry.
	
	The Lagrangian in Eq. (\ref{Lag1}) has no lepton number or equivalently (through sphaleron process) no baryon number violating term, one of the three Sakharov conditions. Thus, to have successful leptogenesis in the above-mentioned framework, which is the aim of this work, we require to extend the fermion sector with one additional fermion gauge singlet $S_L\sim (1,1,0)$ ($S_L \stackrel{\mathcal{C}}{\leftrightarrow}S_L^C $) per generation. The addition of fermion gauge singlets $S_L$ ensures lepton number violation (LNV) by induced Majorana mass terms through the double seesaw mechanism.
	
	\subsection{Double Seesaw and Neutrino Masses}\label{subsec:DS_NeuMass}
	As discussed in \cite{Patra:2023ltl}, addition of sterile neutrinos (fermion gauge singlets) $S_L$ enable to implement the double seesaw mechanism within the manifest LRSM. The relevant interaction Lagrangian for generation of fermion masses is given by:
	\begin{eqnarray}\label{LRDSM}
		\begin{aligned}
			\mathcal{L}_{LRDSM}&= -\mathcal{L}_{M_D}-\mathcal{L}_{M_{RS}}-\mathcal{L}_{M_S}\\
			&=-\sum_{\alpha, \beta} \overline{\nu_{\alpha L}} [M_D]_{\alpha \beta} N_{\beta R}-\sum_{\alpha, \beta} \overline{S_{\alpha L}} [M_{RS}]_{\alpha \beta} N_{\beta R}\\&-\frac{1}{2} \sum_{\alpha, \beta} \overline{S^c_{\alpha R}} [M_{S}]_{\alpha \beta} S_{\beta L} \mbox{+ h.c.}.
		\end{aligned}
	\end{eqnarray}
	Here $\mathcal{L}_{M_D}$ and $\mathcal{L}_{M_{RS}}$ are the Dirac mass terms connecting $\nu_{L}-N_R$ and $N_R-S_L$ respectively. The term $\mathcal{L}_{M_S}$ represents the bare Majorana mass term for sterile neutrinos $S_L$. We emphasize that the Lagrangian in Eq. (\ref{LRDSM}) can not have a bare Majorana mass term for $N_R$ to ensure $U(1)_{B-L}$ gauge invariance. However, if $N_R$ were singlet fermions, the bare Majorana mass term could be added. Such kind of framework has been explored in Ref. \cite{Kang:2006sn}. In Eq. (\ref{LRDSM}), $S^c_{\alpha R} \equiv C(\overline{S_{\alpha L}})^T$, $C$ stands for charge conjugation operation ($C=i\gamma^2\gamma^0$). We note that the Higgs doublet $H_L$ in our model framework is required just for left-right invariance, and it does not participate in SSB. Hence $<H_L^0>\,=0$ and it prevents the mass term connecting $\nu_L-S_R^c$ through the interaction $\sum_{\alpha, \beta} \overline{\ell_{\alpha L}} (Y_{LS})_{\alpha \beta} \widetilde{H_L}
	S^c_{\beta R} \mbox{+ h.c.}$. 
	
	After the scalar fields acquire VEVs and thus lead to SSB, the total $9\times9$ neutral fermion mass matrix in the flavor basis ($\left(\nu_L, N^c_R, S_L \right)$) becomes
	\begin{equation}
		\mathcal{M}_{LRDSM}=
		\left[ 
		\begin{array}{c  c} 
			\begin{array}{c c} 
				{\bf 0} & M_D \\ 
				M^T_D & {\bf 0}
			\end{array} & 
			\begin{array}{c} 
				{\bf 0} \\ M_{RS}
			\end{array} \\ 
			\begin{array}{c c} 
				\,~{\bf 0}  &\quad  M^T_{RS}
			\end{array} & 
			\begin{array}{c} 
				{M_S}
			\end{array} \\
		\end{array} 
		\right]
		\label{eqn:dss1}       
	\end{equation}
	With the assumed hierarchy $|M_D| \ll |M_{RS}| < |M_s|$, we apply double seesaw approximate block diagonalization to $\mathcal{M}_{LRDSM}$. This gives us the mass matrices of light and heavy neutrinos as follows:
	\begin{eqnarray}\label{MajMasses}
			&&m_\nu \cong M_D ({M^T_{RS}})^{-1}  M_S M^T_D M^{-1}_{RS}\text{ (mass matrix of light neutrinos)}  ,\nonumber \\
			&& m_N \equiv M_R \cong -M_{RS} M^{-1}_S M^T_{RS}\text{ (mass matrix of RH neutrinos)}  ,\nonumber \\
			&& m_S \cong M_S \text{ (mass matrix of sterile neutrinos)} \, .
		\end{eqnarray}
		The detailed discussion on double seesaw approximations and derivations of mass matrices expressed in Eq. (\ref{MajMasses}) can be referred from the Ref. \cite{Patra:2023ltl}. For studying thermal unflavored leptogenesis and following the derived mass relations mentioned in Eq. (\ref{MajMasses}), we have estimated the order of magnitudes for various mass parameters in Table \ref{ordermag}; however, one may refer to Table \ref{ordermagAppTab} presented in Appendix \ref{ordermagApp} for a more general perspective of leptogenesis.
	\setlength{\bigstrutjot}{6pt}
	\begin{table}[!h]
		\centering
		\begin{tabular}{|c|c|c||c|c|c|}
			\hline
			$\mathbf{M_D}$ & $\mathbf{M_{RS}}$ & $\mathbf{M_S}$ & $\mathbf{m_\nu}$\textbf{(eV)} &$\mathbf{m_N}$ &$\mathbf{m_S}$ \\
			\hline
			\hline
			$10$  & $10^{14}$  & $10^{15}$ & $0.01$  & $10^{13}$ & $10^{15}$ \\
			\hline
			$10^2$  & $10^{15}$  & $10^{16}$ & $0.1$  & $10^{14}$ & $10^{16}$ \\
			\hline
			$1$ & $10^{13}$ & $10^{14}$\bigstrut[t]\bigstrut[b] & $0.001$ & $10^{12}$ & $10^{14}$\\
			\hline
		\end{tabular}
		\caption{Order of magnitude estimation of various neutrino masses in LRSM with double seesaw mechanism. All masses except the active neutrino masses are in GeV. The numerical values presented in the first row are relevant to our focus within the context of this work.}
		\label{ordermag}
	\end{table}
	\subsection{Gauge boson masses}\label{subsec:gauge_boson_masses}
	Because of left-right symmetry in our framework, the gauge sector comprises extra gauge bosons, $W_R^{\pm}$ and $Z'$. In Ref. \cite{Patra:2023ltl}, gauge boson mass mixing is discussed as per its relevance to neutrinoless double beta decay. In this subsection, we briefly discuss the mass bounds on $W_R^{\pm}$ and $Z'$ gauge bosons from our model framework. The masses of these extra gauge bosons are given as:
	\begin{equation}\label{massWR}
		M_{W_R}\simeq \frac{1}{2} g_Rv_R
	\end{equation} 
	\begin{equation}\label{massZ'}
		M_{Z'}\simeq \frac{\sqrt{g_{BL}^2+g_R^2}}{g_R}M_{W_R}\simeq 1.2M_{W_R}
	\end{equation}
	Here $v_R$ is the VEV acquired by neutral component of Higgs doublet $H_R$, $<H^0_R>\equiv v_R$ and $g_L=g_R\approx0.632$. Through Eq. (\ref{LRDSM}), we have:
	\begin{equation}\label{GBMLag}
			\sum_{\alpha, \beta} \overline{S_{\alpha L}} [M_{RS}]_{\alpha \beta} N_{\beta R} \subset \sum_{\alpha, \beta} \overline{S_{\alpha L}} (Y_{RS})_{\alpha \beta} \widetilde{H_R}^{\dagger} 
			\ell_{\beta R} \mbox{+ h.c.}
	\end{equation}
	Hence,
	\begin{equation}\label{YRS}
		M_{RS}=Y_{RS}.v_R,
	\end{equation}
	where $Y_{RS}$ is the Yukawa coupling. As $m_N$ is dependent on $M_{RS}$ (refer to Eq.~(\ref{MajMasses})), we assume $M_{RS}(=10^{14}\text{ GeV})$ to ensure that $m_N$ remains within the thermal unflavored regime, as indicated in Table \ref{ordermag}, row one. $Y_{RS}$ can take value in the range: $0<Y_{RS}<1$, which when plugged (for $Y_{RS}\sim 1$) in Eq. (\ref{YRS}) may provide an approximate minimum value for $v_R(\approx10^{14}\text{ GeV})$. By putting this minimum value of $v_R$ in Eq. (\ref{massWR}) and (\ref{massZ'}), we obtain lower bounds for the masses of extra heavy bosons as:
	\begin{equation*}
		M_{W_R}\geq 3.2\times 10^{13}\text{ GeV},~~M_{Z'}\geq 3.8\times 10^{13}\text{ GeV}.
	\end{equation*}
		The values are consistent with the current experimental bounds \cite{ATLAS:2018dcj,ATLAS:2019isd,CMS:2018agk,Li:2020wxi,Dekens:2021bro}.
	
	\section{Connecting low-scale and high-scale CP violation}\label{sec:connec_High_Low}
\begin{figure}
		\centering
		\begin{tikzpicture}[line width=1 pt, scale=0.95]
			\draw[color=red,solid] (-7.0,0.0)--(-5.0,0.0);
			\draw[color=blue,dashed] (-5.0,0.0)--(-2.8,1.0);
			\draw[color=red,solid] (-5.0,0.0)--(-2.8,-1.0);
			\node at (-7.3,0.0) {${N_1}$};
			\node [above] at (-3.5,0.8) {${\Phi}$};
			\node [below] at (-3.5,-0.7) {$\ell$};
			\hspace{-0.2in}
			\draw[color=red,solid] (1.0,0.0)--(1.7,0.0);
			\draw[color=red,solid] (2.9,0.0)--(3.6,0.0);
			\draw[color=blue,dashed] (3.6,0.0)--(5.2,1.0);
			\draw[color=red,solid] (3.6,0.0)--(5.2,-1.0);
			\node at (0.7,0.0) {${N_1}$};
			\node [above] at (+3.3,-0.1) {${N_{2,3}}$};
			\node [above] at (4.5,0.6) {${\Phi}$};
			\node [below] at (4.5,-0.5) {$\ell$};
			\node [above] at (2.3,0.5) {${\Phi}$};
			\node [below] at (2.3,-0.55) {$\ell,\bar{\ell}$};
			\draw[color=blue,dashed] (1.7,0) arc (180:0:0.6cm); 
			\draw[color=red,solid](1.7,0) arc (-180:0:0.6cm); 
		\end{tikzpicture}
		\begin{tikzpicture}[line width=1 pt, scale=0.95]
			\draw[color=red,solid] (-7.0,0.0)--(-6.0,0.0);
			\draw[color=red,solid] (-6.0,0.0)--(-4.3,1.0);
			\draw[color=blue,dashed] (-6.0,0.0)--(-4.3,-1.0);
			\draw[color=blue,dashed] (-4.3,1.0)--(-2.8,1.0);
			\draw[color=red,solid] (-4.3,-1.0)--(-4.3,1.0);		
			\draw[color=red,solid] (-4.3,-1.0)--(-2.8,-1.0);
			\node at (-7.3,0.0) {${N_1}$};
			\node at (-3.7,0.0) {${N_{2,3}}$};
			\node [above] at (-2.6,0.65) {${\Phi}$};
			\node [below] at (-2.6,-0.65) {$\ell$};
			\node [above] at (-5.2,-1.2) {${\Phi}$};
			\node [below] at (-5.2,+1.2) {$\ell$};
		\end{tikzpicture}
		\caption{Tree, 1-loop and vertex diagrams for heavy neutrino decay. The asymmetry $\epsilon_1$ results from interference of the 1-loop diagrams with tree level coupling.}	
		\label{FeynDiagCP}
	\end{figure}
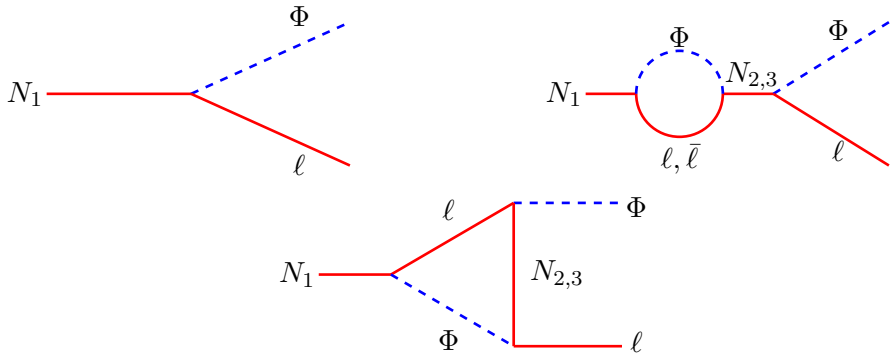

	It is important to establish a connection between the parameters at low-energy (particularly the CP-violating phases), which can be probed in present and future experiments, and at high-energy that would be relevant for leptogenesis. For an unflavored analysis (i.e. considering only $N_1$ decays), this CP-asymmetry ($\epsilon_1$) is defined as
	\begin{equation}\label{epsilon1}
		\epsilon_1=\frac{\Gamma(N_{1}\rightarrow
			\ell\Phi)-\Gamma(N_{1}\rightarrow
			\ell^c\Phi^c)}{\Gamma_{D}}.
	\end{equation}
	Here, $\Gamma_{D}$ is the total decay rate for $N_{1}$ and is expressed as
	\begin{equation}\label{GammaTotal}
		\begin{split}
			\Gamma_{D} & =\Gamma(N_{1}\rightarrow
			\ell\Phi)+\Gamma(N_{1}\rightarrow
			\ell^c\Phi^c)\\
			&=\frac{[Y_D^{\dagger} Y_D ]_{11}m_{N_1}}{8\pi},
		\end{split}
	\end{equation}
	where $Y_D$ is Yukawa coupling matrix as introduced in Eq. (\ref{MDY3}). The asymmetry $\epsilon_1$ arises from the interference between tree and 1-loop wave and vertex diagrams as shown in Fig. \ref{FeynDiagCP}.
	\noindent We work under the assumption that the RH neutrinos have the mass hierarchy as $m_{N_1}<m_{N_2}<m_{N_3}$, so that it is the decays of the $N_1$ that essentially determines the sought-after asymmetry \cite{Covi:1996wh}. 
		\begin{equation}
			\epsilon_1\approx -\frac{3.m_{N_1}}{16\pi(Y_D^\dagger Y_D)_{11}}\left[\frac{Im[(Y_D^\dagger Y_D)^2_{21}]}{m_{N_2}}+\frac{Im[(Y_D^\dagger Y_D)^2_{31}]}{m_{N_3}}\right]
		\end{equation}
		As $M_D=v.Y_D$, we have
		\begin{equation}
			\epsilon_1\approx -\frac{3.m_{N_1}}{16\pi v^2(M_D^\dagger M_D)_{11}}\left[\frac{Im[(M_D^\dagger M_D)^2_{21}]}{m_{N_2}}+\frac{Im[(M_D^\dagger M_D)^2_{31}]}{m_{N_3}}\right].
			\label{epsilon1MD}
	\end{equation}
	Thus, to determine the CP-asymmetry, we need to find the structure of $M_D$ and the masses of right-handed neutrinos ($m_{N_i}$). Many attempts have been made to establish a direct link between the low- and high-scale CP violations, where different assumptions and/or parametrizations of the neutrino Dirac-Yukawa structure ($Y_D$) are made. Among them, the CI parametrization is widely used in which $Y_D$ is expressed in terms of a complex orthogonal matrix $R$ \cite{Casas:2001sr}. Other commonly used parametrizations can be referred from Ref. \cite{Pascoli:2003uh,Branco:2011zb,Joshipura:2001ui,Davidson:2001zk,Rahat:2020mio}. As discussed in detail in Ref. \cite{Branco:2011zb}, such parametrization can lead to other unknown parameters at high-energy, relevant for leptogenesis but inaccessible to experiments. In Ref. \cite{Zhang:2020lir}, the Casas-Ibarra parametrization is used to relate low- and high-energy CP violations through an arbitrary orthogonal matrix $R$, which then is removed by considering discrete left-right symmetry in the framework of MLRSM. In our approach, the theoretical framework of double seesaw in LRSM with considered particle spectrum incorporated with discrete LR symmetry $C$ and the screening condition naturally leads to a direct connection between low- and high-energy CP violations. In what follows, we construct a theoretical structure to have non-zero CP-asymmetry ($\epsilon_1\ne 0$) for leptogenesis and a connection between the low- and high-energy CP violation.
	
	
	\subsection{Screening effect}\label{subsec:screening}
	For the structure of the light neutrino mass matrix, $m_\nu$, to be determined by the structure of the sterile neutrino mass matrix, $m_S$, we apply screening (cancellation) of Dirac structures in the expression of $m_\nu$ of Eq. (\ref{MajMasses}) \cite{Lindner:2005pk}. In reference \cite{Patra:2023ltl}, the authors achieve it by considering $M_D$ and $M_{RS}$ to be proportional to identity ($I$). The consideration was relevant for studying neutrinoless double beta decay. However, we cannot have this screening condition for studying leptogenesis as it leads to vanishing CP-asymmetry ($\epsilon_1=0$). We take the screening condition \cite{Smirnov:2018luj}:
	\begin{equation}\label{screening}
		M_D=\frac{1}{k}M_{RS}^T
	\end{equation}
	Here $k$ is a real constant. It is worth noting that the renormalization group (RG) running of Yukawa couplings ($Y_D$ and $Y_{RS}$) from the LRSM symmetry breaking scale to the EW scale might affect the screening mechanism. While Ref. \cite{Smirnov:2018luj} mentions that RG running would not destroy the screening, one may refer to Ref. \cite{Lindner:2005pk} for a detailed discussion on the impact of RGE effect on the screening in the context of seesaw mechanism and neutrino masses. With Eq.~(\ref{screening}), the relation between light neutrino and sterile neutrino mass matrices $m_\nu$ and $m_S$ becomes $m_S=k^2 m_\nu$. The light neutrino Majorana mass matrix is diagonalized with the Pontecorvo, Maki, Nakagawa, Sakata (PMNS) mixing matrix $U_{PMNS}\equiv U_\nu$:
	\begin{equation}\label{massLight}
		\hat{m}_\nu = U_{\nu}^\dagger m_\nu U_\nu^*= \text{diag}(m_1,m_2,m_3),\ m_i >0\text{ for }i=1,2,3.
	\end{equation}
	In what follows, we will use the standard parametrization of the PMNS matrix \cite{ParticleDataGroup:2018ovx}:
	\begin{eqnarray}\label{UPMNS}
		\begin{aligned}
			U_{\rm {PMNS}}&\equiv U_{\nu}\\
			&=	\begin{pmatrix} c_{13}c_{12}&c_{13}s_{12}&s_{13}e^{-i\delta}\\
				-c_{23}s_{12}-c_{12}s_{13}s_{23}e^{i\delta}&c_{12}c_{23}-s_{12}s_{13}s_{23}e^{i\delta}&s_{23}c_{13}\\
				s_{12}s_{23}-c_{12}c_{23}s_{13}e^{i\delta}&-c_{12}s_{23}-s_{12}s_{13}c_{23}e^{i\delta}&c_{13}c_{23}
			\end{pmatrix}\underbrace{\begin{pmatrix}
					1 &0&0\\
					0 &e^{i\alpha/2}&0\\
					0 &0&e^{i\beta/2}
			\end{pmatrix}}_{\text{Majorana phase matrix}}
		\end{aligned}
	\end{eqnarray}
	where $\delta$ is the Dirac CP phase ($0\leq\delta\leq2\pi$) and $\alpha, \beta$ are the Majorana CP phases ($0\leq\alpha,\beta\leq2\pi$). All the other parameters have their usual meanings. The sterile neutrino Majorana mass matrix $m_S$ is diagonalized by a unitary matrix $U_S$ as $\hat{m}_S=U_S^\dagger m_S U_S^*$, where we have $\hat{m}_S=\text{diag}(m_{S_1},m_{S_2},m_{S_3})$, $m_{S_k}>0,$ $k=1,2,3$. Since $m_S=k^2 m_\nu$, the diagonalization can be done with the same mixing matrix $U_\nu$:
	\begin{equation}\label{USUnu}
		U_S=U_\nu.
	\end{equation}
	So for the considered scenario, the light neutrino masses $m_i$ and the sterile neutrino masses $m_{S_k}$ are related as
	\begin{equation}\label{mimSi}
		m_i=\frac{1}{k^2}m_{S_i},\, i=1,2,3
	\end{equation}
	
	For normal ordering (NO) mass spectrum of active neutrinos ($m_1<m_2<m_3$), we have:
		\begin{equation}\label{NOlightneutrino}
			\begin{aligned}
				& m_1=\text { lightest neutrino mass }\\
				& m_2=\sqrt{m_1^2+\Delta m_{\mathrm{sol}}^2}, \\
				& m_3=\sqrt{m_1^2+\Delta m_{\mathrm{atm}}^2}.
			\end{aligned}
		\end{equation}
		Therefore from Eq. (\ref{mimSi}), the relations between masses of active and sterile neutrinos become:
		\begin{eqnarray}\label{mimSi_NO}
			m_{S_1} = \frac{m_{1}}{m_{3}} m_{S_3}\,, \quad 
			m_{S_2} = \frac{m_{2}}{m_{3}} m_{S_3}\,, \quad  m_{S_1} <  m_{S_2} < m_{S_3}\,. 
		\end{eqnarray}
		For inverted ordering (IO) ($m_3<m_1<m_2$), we have:
		\begin{equation}
			\begin{aligned}
				& m_3=\text { lightest neutrino mass }\\
				& m_1=\sqrt{m_3^2+\Delta m_{\mathrm{atm}}^2},\\
				& m_2=\sqrt{m_3^2+\Delta m_{\mathrm{sol}}^2+\Delta m_{\mathrm{atm}}^2},
			\end{aligned}
		\end{equation}
		and masses relations from Eq. (\ref{mimSi}) become:
		\begin{eqnarray}
			m_{S_1} = \frac{m_{1}}{m_{2}} m_{S_2}\,, \quad
			m_{S_3} = \frac{m_{3}}{m_{2}} m_{S_2}\,, \quad  m_{S_3} <  m_{S_1} < m_{S_2}\,. 
			\label{eqn:MSe2} 
		\end{eqnarray}
		In both orderings, we have $\Delta m_{\mathrm{sol}}^2=\Delta m_{21}^2$ and $\Delta m_{\mathrm{atm}}^2=\left|\Delta m_{31}^2\right|$ \cite{deSalas:2020pgw}.

	\subsection{Choice of basis}\label{subsec:basisChoice}
	We will work in the basis where the charged lepton mass matrix is diagonal. The right-handed neutrino Majorana mass matrix $m_N$ can be diagonalized by a unitary matrix $U_N$ as $\hat{m}_N=U_N^\dagger m_N U_N^*$. Here, we have $\hat{m}_N=diag(m_{N_1},m_{N_2},m_{N_3})$ with $m_{N_i}(i=1,2,3)$ being the mass of the heavy RH Majorana neutrino $N_i$. By using the screening result ($U_S=U_\nu$) from Eq. (\ref{USUnu}), we have 
	\begin{eqnarray}\label{MSinv}
		\begin{aligned}
			&\hat{m}_S=U_\nu^\dagger m_S U_\nu^*\\
			\implies &m_S^{-1}=U_\nu^* \hat{m}_S^{-1}U_\nu^\dagger.
		\end{aligned}
	\end{eqnarray}
	Now using $m_S^{-1}$ from Eq. (\ref{MSinv}) in the expression of $m_N$ from Eq. (\ref{MajMasses}) and noting that $\hat{m}_N=U_N^\dagger m_N U_N^*$, we have:
	\begin{equation}\label{mNdiag}
		\hat{m}_N= -\underbrace{U_N^\dagger M_{RS}U_\nu^*} \hat{m}_S^{-1}\underbrace{U_\nu^\dagger M_{RS}^T U_N^*}
	\end{equation}
	For Eq. (\ref{mNdiag}) to be consistent, the right-hand side should be diagonal. As $\hat{m}_S^{-1}$ is diagonal, it implies that
	\begin{equation}\label{mNimplies}
		U_N^\dagger M_{RS}U_\nu^*=\hat{m}_{RS}.
	\end{equation}
	Since we have considered $\mathcal{C}$ symmetry as the additional discrete symmetry in our model framework, we have $M_D$ and through Eq.~(\ref{screening}), $M_{RS}$ as the symmetric matrices\footnote{This is one of the reasons for considering $\mathcal{C}$ symmetry as the additional discrete LR symmetry in our model framework. }. Thus from Eq.~(\ref{mNimplies}), we must have
	\begin{equation}\label{UNUnu}
		U_N=U_\nu.
	\end{equation}
	Therefore, the diagonalization of right-handed neutrino Majorana mass matrix $m_N$ can be performed with the same mixing matrix $U_\nu$. The screening condition Eq.~(\ref{screening}) also modifies to:
	\begin{equation}\label{screeningNew}
		M_D=\frac{1}{k}M_{RS}
	\end{equation}
	
	\subsection{Determining $M_D$}\label{subsec:MDderivation}
	Determining $M_D$ and hence $Y_D$ is essential for determining the CP-asymmetry required for leptogenesis through the decay of heavy neutrinos. We derive here the matrix structure of $M_D$ analytically and show the connection between low- and high-energy CP violations. 
	
	From Eq. (\ref{MajMasses}), the heavy neutrino mass matrix is
	\begin{equation*}
		m_N=-M_{RS}m_S^{-1}M_{RS}^T.
	\end{equation*} 
	By using $m_S=k^2 m_\nu$ and noting that $M_D, M_{RS}$ are symmetric matrices\footnote{We could not get the final expression of $M_{RS}$ as in Eq. (\ref{MRSfirst}), if it was not symmetric. This is another compelling reason for choosing $\mathcal{C}$ symmetry as the additional discrete LR symmetry in our model-framework.}, we can rewrite it as
	\begin{equation}
		m_N=-\frac{1}{k^2}M_{RS}m_\nu^{-1}M_{RS}.
	\end{equation}
	Simplifying for $M_{RS}$, we get the final expression as:
	\begin{equation}\label{MRSfirst}
		M_{RS}=m_\nu\sqrt{-k^2 m_\nu^{-1}m_N}.
	\end{equation}
	Now by using the Eq. (\ref{UNUnu}) and (\ref{screeningNew}), and extracting the square root of matrices in Eq. (\ref{MRSfirst}), we get the expression for $M_D$:
	\begin{equation}\label{MDexpression}
		M_D=\frac{1}{k}M_{RS}=i.U_\nu\hat{m}_\nu(\hat{m}_\nu^{-1}\hat{m}_N)^{1/2}U_\nu^T
	\end{equation}
	The full derivation of $M_D$ obtained in Eq.~(\ref{MDexpression}) is given in Appendix \ref{AppendixMD}. We have thus connected low- and high-scale CP violations through Eq.~(\ref{MDexpression}), as $M_D$ is expressed in terms of $U_\nu\equiv U_{PMNS}$. The masses of heavy neutrinos are denoted by $m_{N_i}$ ($i=1,2,3$), with $m_{N_1}<m_{N_2}<m_{N_3}$. Using these in Eq. (\ref{MDexpression}),we have matrix structure of $M_D$ as follows:
		\begin{equation}\label{MDMatrix}
			M_D=i.U_\nu \begin{bmatrix}
				\sqrt{m_1.m_{N_1}} &0 &0\\
				0&\sqrt{m_2.m_{N_2}} &0\\
				0&0 &\sqrt{m_3.m_{N_3}}
			\end{bmatrix} U_\nu^T
	\end{equation}

	\section{Leptogenesis}\label{sec:lepto}
	
	The scenario of leptogenesis to produce an asymmetry requires fulfillment of three Sakharov conditions~\cite{Sakharov:1967dj} as the minimum necessary criteria. Thus, in this section, we present an implementation of thermal leptogenesis in the context of our framework by ensuring the sanctity of the required Sakharov criteria and also by producing the adequate lepton asymmetry required for meeting the observational evidence of BAU from the Planck~\cite{Planck:2018vyg} data. In the most general sense, the Lagrangian terms relevant for leptogenesis in our framework are given by:
	\begin{eqnarray}
		\mathcal{L}_{LG} &=& \sum_{\alpha, \beta} \overline{\nu_{\alpha L}} [M_D]_{\alpha \beta} N_{\beta R}+\overline{S_{\alpha L}} [M_{RS}]_{\alpha \beta} N_{\beta R} \mbox{ + h.c.} \nonumber \\
		&&\subset \sum_{\alpha, \beta} \overline{\ell_{\alpha L}} \left( (Y_\ell)_{\alpha \beta} \Phi + (\widetilde{Y}_\ell)_{\alpha \beta} \tilde{\Phi} \right)
		\ell_{\beta R}+ \overline{S_{\alpha L}} (Y_{RS})_{\alpha \beta} \widetilde{H_R}^{\dagger}
		\ell_{\beta R} \mbox{+ h.c.}
		\label{eq:lg1}
	\end{eqnarray}
	
	Here, the interaction between $\nu_L$ and $N_R$ present in the term with Dirac mass, $M_D$ is responsible for creating an asymmetric production of active neutrinos. The heavy right-handed neutrinos denoted by $N_R$ are unstable, and thus their early decay before electroweak symmetry breaking in the Universe creates the asymmetry. The complex nature of Yukawa coupling matrices $(Y_D)$ ensures that the decays are CP-violating. To achieve out-of-equlibrium conditions, the mass of $N_R$ is taken to be large so that the temperature of the Universe at the time of their decay is less than their rest mass, and hence the probability of inverse interactions decreases, keeping the decays out-of thermal equilibrium. The construct of double seesaw ensures a lepton number and $B+L$ violating interaction, which via the sphelaron process, transfers the created asymmetry into the baryon sector. The second term on the right in Eq.~(\ref{eq:lg1}), is responsible for the decay of $S_L$ into $N_R$ (as the sterile fermions $S_L$ are much heavier than $N_R$) and thus altering the abundance of $N_R$ throughout the cosmological timescale, indirectly affecting leptogenesis. Now, we discuss the generation of asymmetry in the decays of $N_R$ in more detail. For the purpose of this work, our primary emphasis lies in examining the correlation between CP asymmetry and the Dirac CP phase ($\delta$). Consequently, we shall presume the Majorana phases to be set at zero. However, to ensure comprehensiveness, we will briefly explore the impact of non-zero Majorana phases on the overall scenario whenever necessary.

	\begin{table}[]
		\centering
		\begin{tabular}{|c|ccc|}
			\hline Parameter & best-fit $\pm 1 \sigma$ & $2 \sigma$ range & $3 \sigma$ range \\
			\hline 
			&&&\\[-8pt]
			$\Delta m_{21}^2\left[10^{-5} \mathrm{eV}^2\right]$ & $7.50_{-0.20}^{+0.22}$ & $7.12-7.93$ & $6.94-8.14$ \\[5pt]
			$\left|\Delta m_{31}^2\right|\left[10^{-3} \mathrm{eV}^2\right](\mathrm{NO})$ & $2.55_{-0.03}^{+0.02}$ & $2.49-2.60$ & $2.47-2.63$ \\[5pt]
			$\left|\Delta m_{31}^2\right|\left[10^{-3} \mathrm{eV}^2\right](\mathrm{IO})$ & $2.45_{-0.03}^{+0.02}$ & $2.39-2.50$ & $2.37-2.53$ \\[5pt]
			$\sin ^2 \theta_{12} / 10^{-1}$ & $3.18 \pm 0.16$ & $2.86-3.52$ & $2.71-3.69$ \\
			$\theta_{12} /^{\circ}$ & $34.3 \pm 1.0$ & $32.3-36.4$ & $31.4-37.4$ \\[5pt]
			$\sin ^2 \theta_{23} / 10^{-1}(\mathrm{NO})$ & $5.74 \pm 0.14$ & $5.41-5.99$ & $4.34-6.10$ \\[5pt]
			$\theta_{23} /^{\circ}(\mathrm{NO})$ & $49.26 \pm 0.79$ & $47.37-50.71$ & $41.20-51.33$ \\[5pt]
			$\sin ^2 \theta_{23} / 10^{-1}(\mathrm{IO})$ & $5.78_{-0.17}^{+0.10}$ & $5.41-5.98$ & $4.33-6.08$ \\[5pt]
			$\theta_{23} /{ }^{\circ}(\mathrm{IO})$ & $49.46_{-0.97}^{+0.60}$ & $47.35-50.67$ & $41.16-51.25$ \\[5pt]
			$\sin ^2 \theta_{13} / 10^{-2}(\mathrm{NO})$ & $2.200_{-0.062}^{+0.069}$ & $2.069-2.337$ & $2.000-2.405$ \\[5pt]
			$\theta_{13} /^{\circ}(\mathrm{NO})$ & $8.53_{-0.12}^{+0.13}$ & $8.27-8.79$ & $8.13-8.92$ \\[5pt]
			$\sin ^2 \theta_{13} / 10^{-2}(\mathrm{IO})$ & $2.225_{-0.070}^{+0.064}$ & $2.086-2.356$ & $2.018-2.424$ \\[5pt]
			$\theta_{13} /{ }^{\circ}(\mathrm{IO})$ & $8.58_{-0.14}^{+0.12}$ & $8.30-8.83$ & $8.17-8.96$ \\[5pt]
			$\delta / \pi(\mathrm{NO})$ & $1.08_{-0.12}^{+0.13}$ & $0.84-1.42$ & $0.71-1.99$ \\[5pt]
			$\delta / \pi(\mathrm{IO})$ & $1.58_{-0.16}^{+0.15}$ & $1.26-1.85$ & $1.11-1.96$ \\[5pt]
			\hline
		\end{tabular}
		\caption{The current updated estimates of experimental values of
				neutrino oscillation parameters for global best-fits and for $1\sigma$ to $3\sigma$ ranges taken from~\cite{deSalas:2020pgw}.}
		\label{tab:exp}
	\end{table}
	
	\subsection{CP asymmetry and out-of-equilibrium dynamics}
	This subsection discusses the conditions required for effective out-of-equilibrium decays of heavy neutrinos and the generation of required CP asymmetry. It is known that the interaction rates which are equal to or slower in order than the Hubble expansion rate, are not fast enough to equilibriate particle distributions. This roughly translates to the following inequality,
	\begin{equation}
			\Gamma_D \le H (T=m_{N_1}).
			\label{eq:dr}
	\end{equation}
	\begin{table}[]
		\centering
		\begin{tabular}{|c|*{3}{>{\centering\arraybackslash}p{2.2cm}|}*{1}{>{\centering\arraybackslash}p{1.9cm}|}*{1}{>{\centering\arraybackslash}p{2.2cm}|}}
			\hline
			\textbf{NO/IO}&$\mathbf{m_\nu}$(eV) & $\mathbf{m_{N}}$(GeV) & $\mathbf{({M_D}^{\dagger}M_D)_{11}}$ & $\Gamma_{D}$ & $H(T=m_{N_1})$  \\
			\hline
			\multirow[c]{3}{*}[-3pt]{\rotatebox[origin=c]{90}{\textbf{Normal Ordering (NO)}}}&$~~$ $\{0.010$,  $~~~$ $~~~~0.013$,$~~~$ $0.051\}$  & $\{1\times10^{10}$, $1\times 10^{11}$, $5\times 10^{11}\}$ & $\mathbf{~~~~~~~~~~~~~~~~}$ $1.04$ & $\mathbf{~~~~~~~~~}$ $6855.27$ & $\mathbf{~~~~~~~~}$ $143.76$ \\
			\cline{2-6}
			&$\mathbf{\{0.010}$, $\mathbf{0.013}$, $\mathbf{0.051\}}$  & $\mathbf{\{1\times10^{13}}$, $\mathbf{1\times10^{14}}$, $\mathbf{5\times 10^{14}\}}$ & $\mathbf{~~~~~~~}$ $\mathbf{1042.64}$ & $\mathbf{~~~}$ $\mathbf{6.86\times10^{9}}$ & $\mathbf{~~~~~}$ $\mathbf{1.44\times10^{8}}$ \\
			\cline{2-6}
			&$~~$ $\{0.010$,  $~~~$ $~~~~0.013$,$~~~$ $0.051\}$   & $\{1\times10^7$, $1\times 10^{8}$, $5\times 10^{8}\}$ & $\mathbf{~~~~~}$ $1.04\times10^{-3}$ & $\mathbf{~~~~~~~~~~~~}$ $6.86\times10^{-3}$ & $\mathbf{~~~~~}$ $1.44\times10^{-4}$ \\
			\hline\hline
			\multirow[c]{3}{*}[-4pt]{\rotatebox[origin=c]{90}{\textbf{Inverted Ordering (IO)}}}&	$~~$ $\{0.050$,  $~~~$ $~~~~0.051$,$~~~$ $0.010\}$  & $\{1\times10^{10}$, $1\times 10^{11}$, $5\times 10^{11}\}$ & $\mathbf{~~~~~~~~~~~~}$ $2.00$ & $\mathbf{~~~~~~~~~}$ $13169.70$ & $\mathbf{~~~~~~~~}$ $143.76$ \\
			\cline{2-6}
			&$\mathbf{\{0.050}$, $\mathbf{0.051}$, $\mathbf{0.010\}}$  & $\mathbf{\{1\times10^{13}}$, $\mathbf{1\times10^{14}}$, $\mathbf{5\times 10^{14}\}}$ & $\mathbf{~~~~~~~}$ $\mathbf{2003.03}$ & $\mathbf{~~~~~}$ $\mathbf{1.32\times10^{10}}$ & $\mathbf{~~~~~}$ $\mathbf{1.44\times10^{8}}$ \\
			\cline{2-6}
			&$~~$ $\{0.050$,  $~~~$ $~~~~0.051$,$~~~$ $0.010\}$  & $\{1\times10^7$, $1\times 10^{8}$, $5\times 10^{8}\}$ & $\mathbf{~~~~~}$ $2.00\times10^{-3}$ & $\mathbf{~~~~~~~}$ $1.32\times10^{-2}$ & $\mathbf{~~~~~}$ $1.44\times10^{-4}$ \\
			\hline
		\end{tabular}
		\caption{Table represents the obtained values of total decay rate for lightest right-handed neutrino $(\Gamma_D)$ and the Hubble's expansion rate $(H)$ at $T=m_{N_1}$ for various combination of heavy and light neutrino masses. The second row for each mass ordering in bold represents the benchmark points for our analysis.}
		\label{tab:outofeq}
	\end{table}
	Here, $\Gamma_D$ is the total decay rate for lightest right-handed neutrino $(N_1)$ and is expressed in Eq.~(\ref{GammaTotal}) and $H(T=m_{N_1})$ is the Hubble's expansion rate at the time when Universe temperature equals the mass of $N_1$. Upon checking the viability of Eq.~(\ref{eq:dr}) for a particular parameter choice, one may infer that if indeed $\Gamma_D<H$, then this is referred to as being in \textquotedblleft weak washout\textquotedblright~regime and here leptogenesis is almost unavoidable in the sense that all three Sakharov conditions have been fulfilled. Conversely, it is interesting to note that $\Gamma_D>H$ does not ensure an unsuccessful leptogenesis but rather a more detailed analysis of the interplay between decays, inverse decays, and scatterings processes via a complete Boltzmann treatment decide the final asymmetry value in this case~\cite{Davidson:2008bu, Joshipura:2001ya}. Such a regime is the focus of our work and is referred to as \textquotedblleft strong washout\textquotedblright~regime. Using the oscillation data from Table~\ref{tab:exp}, we have presented the values of $\Gamma_D$ and $H$ for 3 different choices of input parameters in Table~\ref{tab:outofeq}.
	
	The out-of-equilibrium decays of the heavy Majorana neutrinos involve CP-asymmetric amplitudes if the neutrino Yukawa couplings $Y_{D}$ are allowed to be complex. As given in Eq.~(\ref{epsilon1MD}), the generalized analytical expression for asymmetry parameter is given as
	\begin{equation}
		\epsilon_1\approx -\frac{3.m_{N_1}}{16\pi v^2(M_D^\dagger M_D)_{11}}\left[\frac{\text{Im}[(M_D^\dagger M_D)^2_{21}]}{m_{N_2}}+\frac{\text{Im}[(M_D^\dagger M_D)^2_{31}]}{m_{N_3}}\right].
		\label{eq:cp1}
	\end{equation}
	To highlight the connection between this asymmetry parameter, $\epsilon_1$ (relevant for leptogenesis) and the CP-violating phase, $\delta$ (coming from low-scale physics), we derive the various terms in Eq.~(\ref{eq:cp1}) analytically in terms of the CP-violating phase, using the structure of $M_D$ given in Eq.~(\ref{MDMatrix}) along with the values of experimental parameters given in Table~\ref{tab:exp} and by fixing rest of the input parameters. Now, given the two set of possible mass hierarchy for active neutrinos~(NO \& IO), we present our analysis separately for the two cases, in the subsequent text.

		\begin{figure}[]
			\centering
			\hspace{-0.2 in}\includegraphics[width=0.95\textwidth]{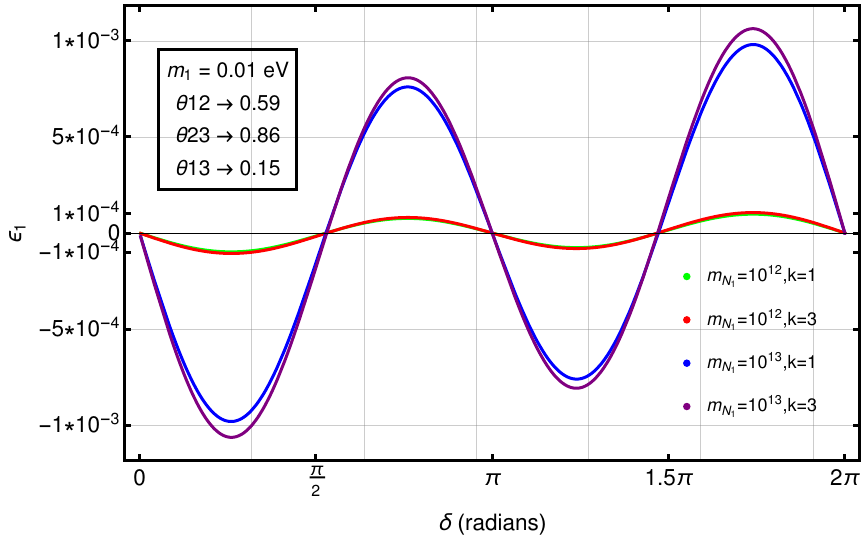}
			\caption{\textbf{(NO case)} Plot for the dependence of $\epsilon_1$ on CP-violating Dirac phase,~$\delta$ for different combinations of right-handed neutrino mass and the hierarchy in that sector. Value of variable $k$ represents here the masses of heavier right-handed neutrino for the structure: $M_{N_2}=1\times10^{k}\times M_{N_1}$ and $M_{N_3}=5\times10^{k}\times M_{N_1}$.
			}
			\label{fig:3}
		\end{figure}
		\subsubsection{Normal Ordering}
		\label{subsec:NO1}
		From Eq.~(\ref{eq:cp1}), for normal ordering (NO) mass spectrum of active neutrinos, we have:
       \begin{equation}
		\begin{split}
			& \text{Im}[(M_D^\dagger M_D)^2_{21}]=  \sin{\delta}\left(2.06+16\cos{\delta}\right)\times10^{6}\text{~GeV}^2\\
			& \text{Im}[(M_D^\dagger M_D)^2_{31}]=  \sin{\delta}(-2.06+11.86\cos{\delta})\times10^{6}\text{~GeV}^2\\
			& (M_D^\dagger M_D)_{11}=  1042.64 \text{~GeV}^2
			\label{eq:analasymm}
		\end{split}
	\end{equation}
	Here, we have used the benchmark point values $m_{N_1}=10^{13}\text{ GeV}$, $m_{N_2}=1\times10^{14}\text{ GeV}$, $m_{N_3}=5\times10^{14}\text{ GeV}$ and $m_{1}=0.01\text{ eV}$ of input parameters from the Table~\ref{tab:tabbp}~(Row 1 in NO) for the calculations. In Fig.~\ref{fig:3}, we plot the variation of $\epsilon_1$ against the allowed range of CP-violating phase $\delta~[0,2\pi]$. From the plot, the dependence of $\epsilon_1$ on this Dirac CP phase is clearly evident, and the sinusoidal nature of this dependence, as mentioned in Eq.~(\ref{eq:analasymm}) is visible from the vanishing asymmetry value at angles 0, $\pi$ and $2\pi$. Also, one sees a sign change of asymmetry parameter at regular intervals, which restricts the allowed parameter space from the requirements of the final baryon asymmetry, as discussed later in subsection \ref{subsec:CEA}.
	
	\begin{figure}
		\centering
		\begin{subfigure}[b]{0.75\textwidth}
			\includegraphics[width=1\linewidth]{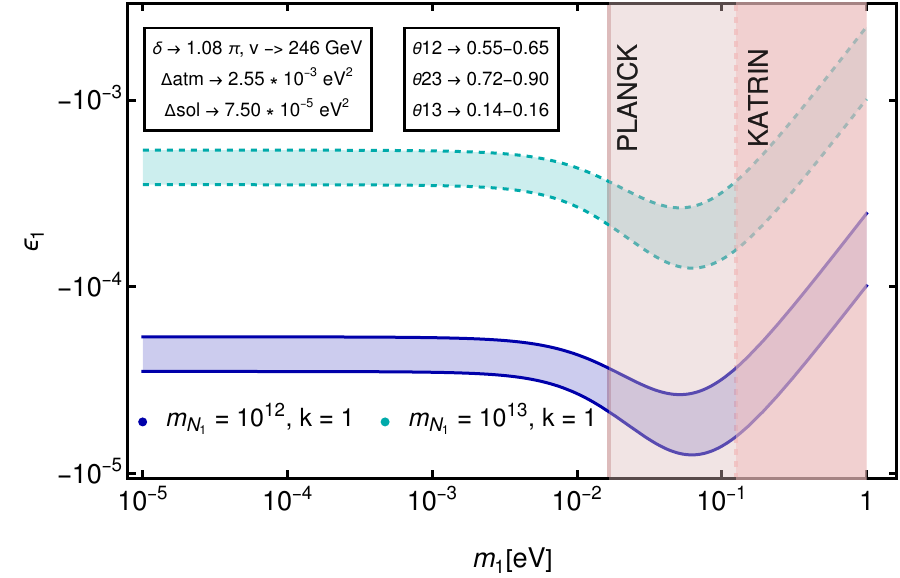}
			\caption{}
			\label{fig:Ng1} 
		\end{subfigure}
		
		\begin{subfigure}[b]{0.75\textwidth}
			\includegraphics[width=1\linewidth]{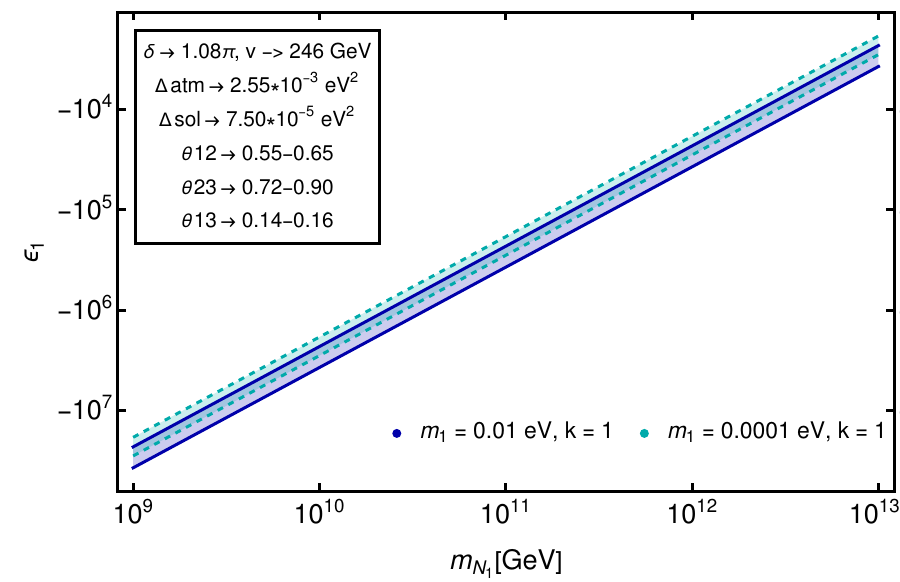}
			\caption{}
			\label{fig:Ng2}
		\end{subfigure}
		
		\caption[]{\textbf{(NO case)} The above figure shows two region plots depicting the dependence of asymmetry parameter,~$\epsilon_1$ on lightest active neutrino mass,~$m_1$ in plot (a) and on lightest right-handed neutrino mass,~$m_{N_1}$ in plot (b). For both the plots we vary the values of oscillation parameters $\theta_{12}$, $\theta_{23}$ and $\theta_{13}$ within their allowed $3\sigma$ range. CP phase $\delta$ has been set to its best-fit value of $1.08\pi$. Value of variable $k$ represents here the masses of heavier right-handed neutrino for the structure: $M_{N_2}=1\times10^{k}\times M_{N_1}$ and $M_{N_3}=5\times10^{k}\times M_{N_1}$. The vertical pink bands in plot~(a) represent the bound corresponding to the upper limit on the sum of light neutrino masses of $0.12~\text{eV}$ reported by the Planck~\cite{Planck:2018vyg} and the prospective bound of $0.20~\text{eV}$ that can be set by the KATRIN~\cite{KATRIN:2019yun} collaboration.}
		\label{fig:4}
	\end{figure}
	
	Using the benchmark values of all the input parameters along with the best-fit value of $\delta=1.08\pi$~\cite{deSalas:2020pgw}, and the standard Higgs VEV of $v\simeq 246 \text{ GeV}$, the asymmetry parameter here is numerically obtained as:
	\begin{equation}
		\epsilon_1 \sim -3.8\times 10^{-4}.
		\label{eq:asymNO}
	\end{equation}
	Also, by using all the relevant parameters mentioned above, the generic structure of $M_D$ as given in Eq. (\ref{MDMatrix}) is calculated numerically to be equal to,
	\begin{equation}
		M_D= \begin{pmatrix}
			~~~~1.73 + 20.90i & ~-4.97 - 7.57i & ~-4.28 - 22.42i\\
			-4.97 - 7.57i & ~~~~-0.56 + 104.04i & ~-0.03 + 63.52i\\
			~-4.28 - 22.42i & ~~~-0.03 + 63.52i & ~~~~~0.36 + 81.37i
		\end{pmatrix}
		\label{eq:cpmat}
	\end{equation}
	From Eq.~(\ref{eq:cpmat}), we have Yukawa matrix as $Y_D=M_D/v$. The elements of $Y_D$ matrix are used to calculate the various interaction rates required for solving the Boltzmann equations (\ref{be1}) and~(\ref{be2}).
	
	We now plot the dependence of CP asymmetry parameter $\epsilon_1$ on the lightest active and right-handed neutrino masses, respectively, in figures \ref{fig:Ng1} and~\ref{fig:Ng2}. For both the plots, we vary the values of oscillation parameters $\theta_{12}$, $\theta_{23}$ and $\theta_{13}$ within their allowed $3\sigma$ range and thus obtain bands of calculated $\epsilon_1$ values concerning a range of $m_1$ and $m_{N_1}$. In Fig.~\ref{fig:Ng1}, we also depict the upper bound for the sum of light neutrino masses obtained through Planck~\cite{Planck:2018vyg} and KATRIN~\cite{KATRIN:2019yun} collaborations. Thus, within the allowed range of $m_1$ smaller than $0.12~\text{eV}$, we see that for most of the parameter space, the asymmetry parameter does not depend on $m_1$. The behaviour of the plot changes slightly at higher mass values near to $10^{-2}~\text{eV}$. In Fig.~\ref{fig:Ng2}, we see that $\epsilon_1$ directly depends on the value of $m_{N_1}$ for the entire range of interest. It is also evident from this plot that for a given $m_{N_1}$, the value of $\epsilon_1$ is barely sensitive to the choice of $m_1$. The values of other relevant input parameters are given within the plots, and the CP-violating phase is set to its best-fit value of $\delta=1.08\pi$ for the NO case.
	
\begin{figure*}[!ht]
		\includegraphics[width=0.44\textwidth]{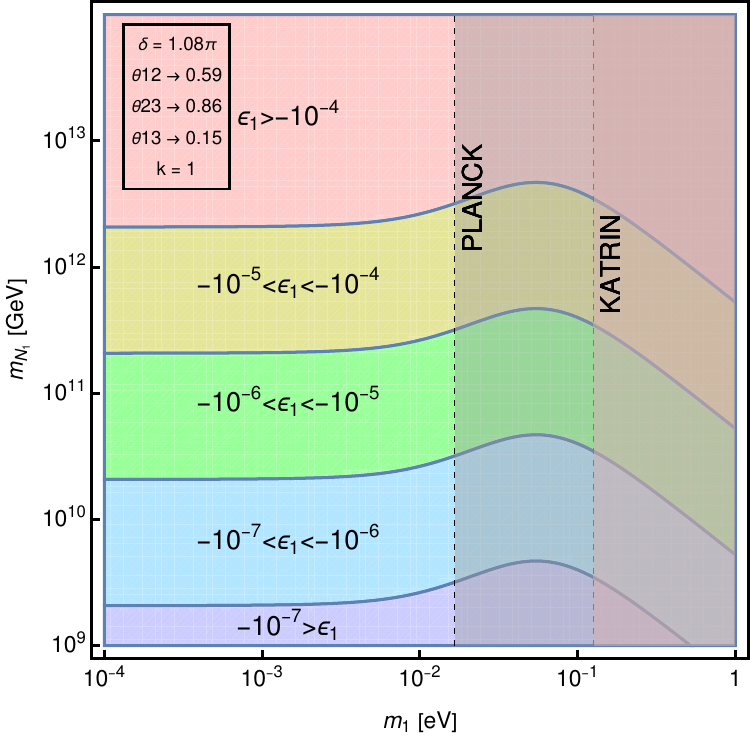}
		\includegraphics[width=0.56\textwidth]{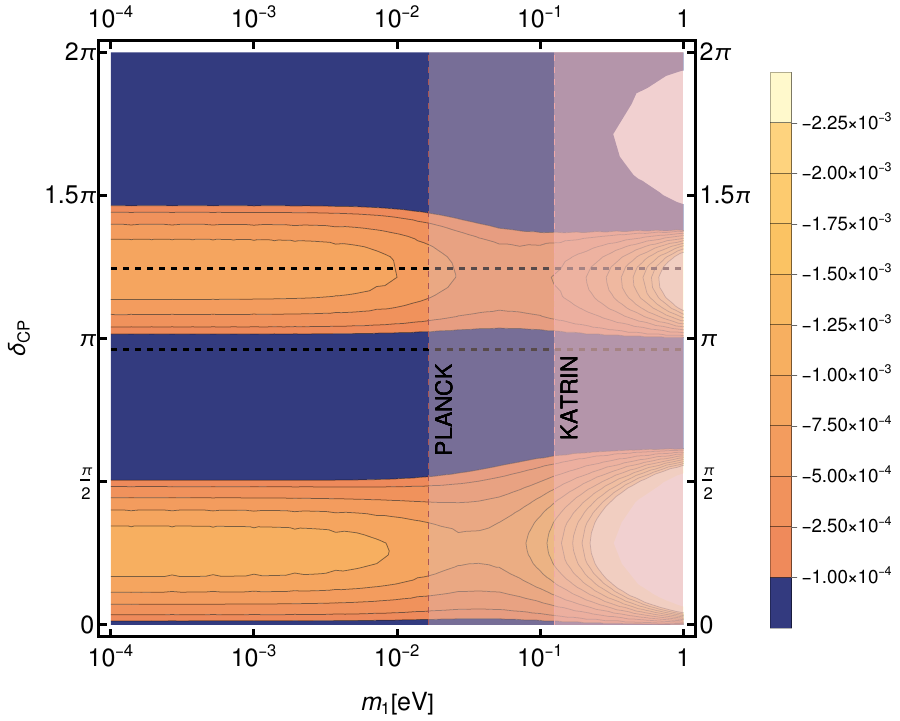}
		\caption{\textbf{(NO case)} The region plots here explore the dependence of asymmetry parameter $(\epsilon_1)$ on a set of other input parameters. In the left plot, input variables are the masses of lightest active and right-handed neutrinos, respectively on the x and y axis. The values of other relevant parameters are mentioned within the plot. In the right plot, input variables are the mass of lightest active neutrino and the value of Dirac CP phase, respectively on x and y axis. All other relevant parameters are fixed at their best-fit values given in Table~\ref{tab:exp}. The masses of right-handed neutrinos are set at their benchmark values of $m_{N_1}=1\times10^{13}~\text{GeV}$, $m_{N_2}=1\times10^{14}~\text{GeV}$ and $m_{N_3}=5\times10^{14}~\text{GeV}$. For both the plots, we also show the experimental bounds on lightest active neutrino mass from Planck and KATRIN experiments marked by vertical bands. Also, in the right plot, the black dashed lines running horizontally marks the currently accepted $1\sigma$ band for Dirac CP phase ($\delta_{CP}$)~\cite{deSalas:2020pgw}.}
		\label{fig:9}
	\end{figure*}

	In Fig.~\ref{fig:9}, we show region plots depicting the dependence of $\epsilon_1$ in $m_1-m_{N_1}$ plane and $m_1-\delta_{\text{CP}}$ plane, respectively in left and right plots. Bands of different $\epsilon_1$ values are obtained for the combination of input parameter values. A direct proportionality of $\epsilon_1$ value on $m_{N_1}$ and almost negligible dependence on $m_1$ for the allowed region of parameter space is clearly visible in the left plot. We also see that for the given range of input variables, value of $\epsilon_1$ varies from $-10^{-7}\lesssim\epsilon_1\lesssim-10^{-4}$. In the right plot, we find interesting contours for different values of $\epsilon_1$ ranging from $-2.5\times10^{-4}$ to $-1.0\times10^{-4}$. Such features directly result from the sinusoidal dependence of $\delta$ parameter. The blue contours in the right plot represents the set of $\delta$ and $m_1$ values for which we obtain a positive value of $\epsilon_1$. In the case of thermal unflavored (vanilla) leptogenesis, it can be seen from equations~(\ref{eq:Yanal}) and~(\ref{eqnarr:asym}) that a positive value of $\epsilon_1$ leads to an obtained baryon asymmetry value with wrong sign~(i.e. negative sign). Thus, our interest lies in the parameter space where we obtain a negative value of $\epsilon_1$.\\

		\begin{figure}
		\centering
		\begin{subfigure}[b]{0.95\textwidth}
			\includegraphics[width=1\linewidth]{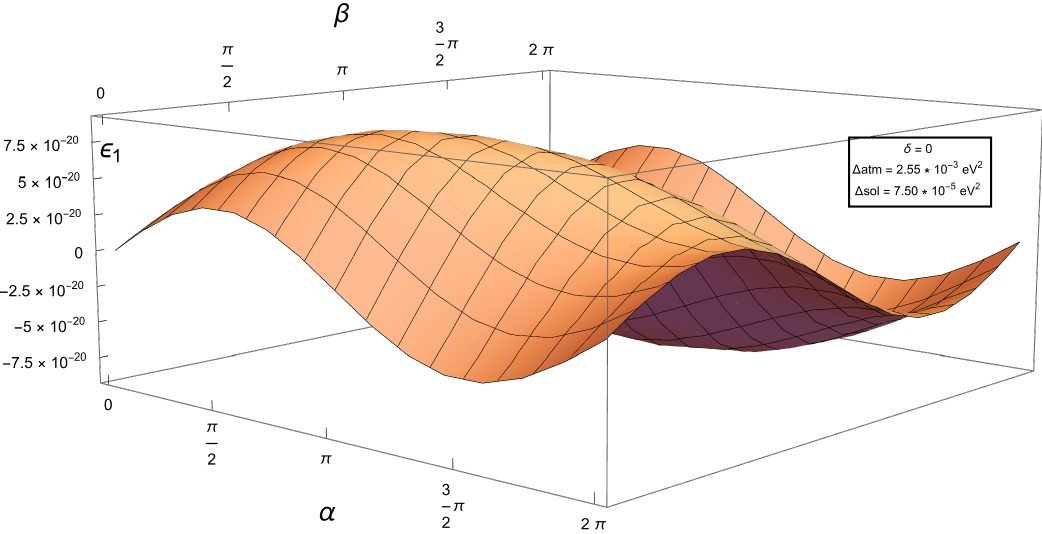}
			\caption{}
			\label{fig:NOMPA} 
		\end{subfigure}
		
		\begin{subfigure}[b]{0.95\textwidth}
			\includegraphics[width=1\linewidth]{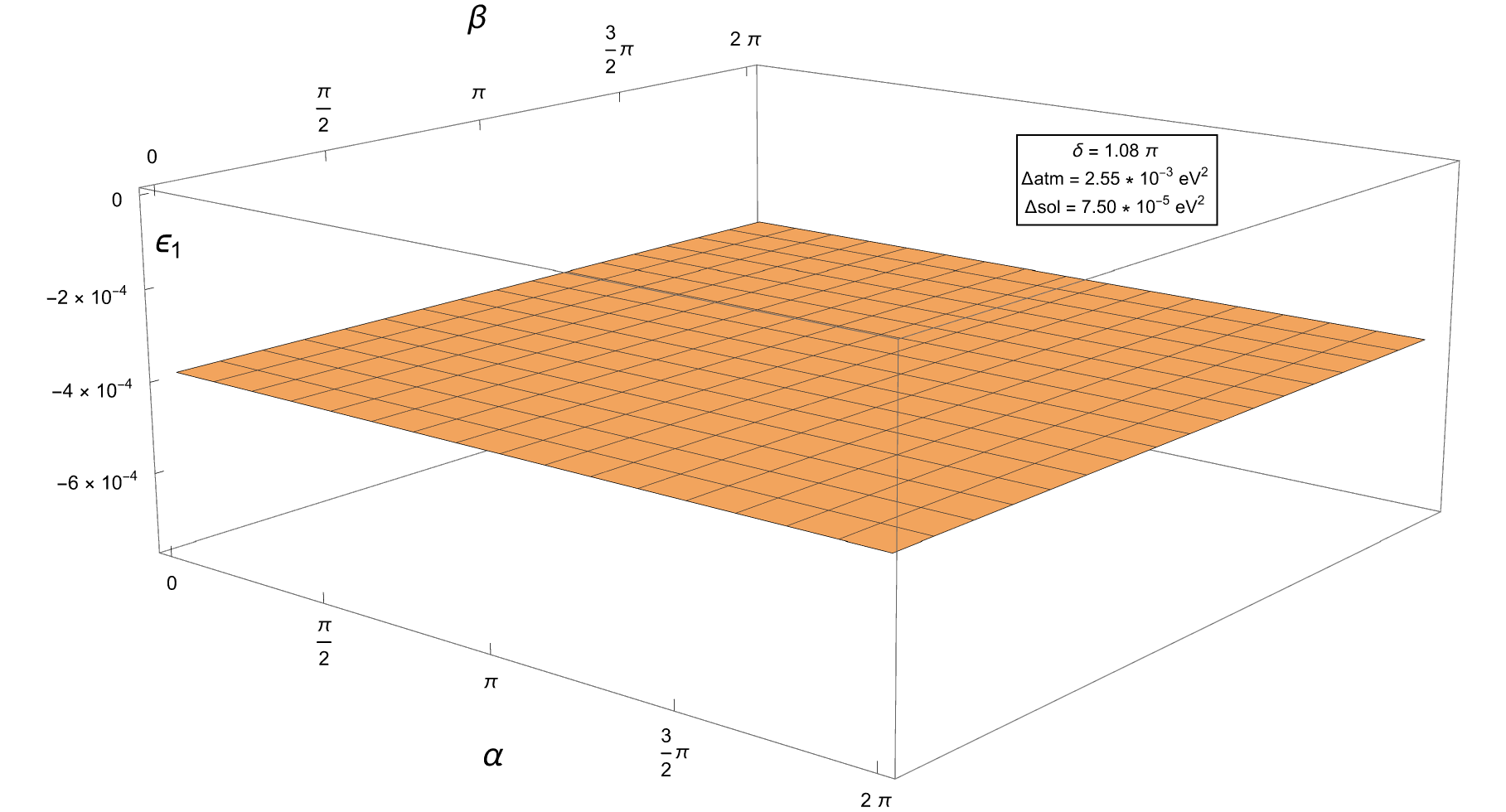}
			\caption{}
			\label{fig:NOMPB}
		\end{subfigure}
		
		\caption{\textbf{(NO case)} The above figure shows two 3D plots (a) and (b) depicting the dependence of asymmetry parameter~$\epsilon_1$ on both the Majorana phases $\alpha$ and $\beta$ for the case of zero~$(\delta=0)$ and best-fit~$(\delta=1.08\pi)$ Dirac phase, respectively. The values of $\alpha$ and $\beta$ run from $0$ to $2\pi$ for both the plots. In plot (a), the Dirac CP phase is set at zero and thus we obtain a $(\alpha, \beta)$ dependent region of $\epsilon_1$ values~(light copper colored region). In plot (b), we see that there is almost no dependence of $\epsilon_1$ on the Majorana phases and thus a flat plane~(orange colored region) is obtained for the entire plot range. The values of all the other relevant input parameters are mentioned within these plots. The range of obtained $\epsilon_1$ value along with its sign are shown on the z-axis in these plots.}
		
		\label{fig:NOMP}
	\end{figure}
	\underline{\textbf{Inclusion of Majorana Phases}}\\[.1in]
	Our work primarily focuses on the Dirac CP phase ($\delta$) as the exclusive source of CP violation for phenomenological discussions. However, here, we also incorporate the Majorana phases ($\alpha, \beta$) introduced in Eq. (\ref{UPMNS}) to encompass all aspects of CP violation and to explore the possibility of lowering the scale of the leptogenesis scenario. Thus, we express the various terms in Eq. (\ref{eq:cp1}) analytically in terms of the Majorana phases for two different values of the Dirac phase: \textbf{(1.)} $\delta=0$ (CP conserving) and \textbf{(2.)} $\delta=1.08\pi$ (best-fit).
	\begin{enumerate}
		\item[ \textbf{(1.)}] For $\delta=0$.
		\begin{equation}
			\begin{split}
				& \text{Im}[(M_D^\dagger M_D)^2_{21}]\simeq  -[5.45\sin{\beta}+85.81\sin{\alpha}]\times10^{-10}\text{~GeV}^2+f(\alpha,\beta)\,\mathcal{O}(10^{-26})\\
				& \text{Im}[(M_D^\dagger M_D)^2_{31}]\simeq  [1.43\sin{\alpha}-5.72\sin{\beta}-0.16\sin{(\alpha-\beta)}-2.29\sin{(\alpha+\beta)}]\\
				&\hspace{1.19in}\times10^{-10}\text{~GeV}^2+f(\alpha,\beta)\,\mathcal{O}(10^{-26})\\
				& (M_D^\dagger M_D)_{11}=  [1042.64-[ 2.84\cos{\alpha}-11.37\cos{(\alpha-\beta)}+2.84\cos{\beta}]\times10^{-14}]\text{~GeV}^2
			\end{split}
		\end{equation}
		Here, $f(\alpha,\beta)\mathcal{O}(10^n)$ corresponds to the sine and cosine functions of Majorana phases ($\alpha,\beta$) with an order of magnitude $n$ or less, where $n$ is an integer. Using the benchmark values of all the input parameters and the standard Higgs VEV of $v\simeq 246 \text{ GeV}$, the asymmetry parameter here is numerically obtained as:
		\begin{equation}
			\epsilon_1 \sim [(0.59+3.46\cos{\beta})\sin{\alpha}+(6.23-2.60\cos{\alpha})\sin{\beta}]\times 10^{-20}.
		\end{equation}
		
		\item[ \textbf{(2.)}] For $\delta=1.08\pi$.
		\begin{equation}
			\begin{split}
				& \text{Im}[(M_D^\dagger M_D)^2_{21}]\simeq  3.34\times10^{6}\text{~GeV}^2+f(\alpha,\beta)\,\mathcal{O}(10^{-9})\\
				& \text{Im}[(M_D^\dagger M_D)^2_{31}]\simeq  3.37\times10^{6}\text{~GeV}^2+f(\alpha,\beta)\,\mathcal{O}(10^{-9})\\
				& (M_D^\dagger M_D)_{11}\simeq  1042.64 \text{~GeV}^2+f(\alpha,\beta)\,\mathcal{O}(10^{-14}) 
			\end{split}
		\end{equation}
		With the benchmark values, the asymmetry parameter is numerically obtained as:
		\begin{equation}
			\epsilon_1 \sim -3.8\times 10^{-4}+f(\alpha,\beta)\,\mathcal{O}(10^{-13})
		\end{equation}
	\end{enumerate}
	In Fig. \ref{fig:NOMP}, we depict the variation of the CP asymmetry parameter ($\epsilon_1$) with respect to the Majorana phases $\alpha$ and $\beta$, while holding the Dirac phase ($\delta$) fixed at a CP-conserving value ($\delta=0$) in Fig.~\ref{fig:NOMPA} and at the best-fit value ($\delta=1.08\pi$) in Fig.~\ref{fig:NOMPB}. These plots reveal that the dependence of $\epsilon_1$ on the Majorana phases is several orders of magnitude smaller than its dependence on the Dirac CP phase. Consequently, Majorana phases play an insignificant role in generating the required CP asymmetry.

	\begin{table}[h!]
		\centering
		\begin{tabular}{|*{6}{>{\centering\arraybackslash}p{2.1cm}|}}
			\hline
			\multicolumn{6}{|c|}{\textbf{Normal Ordering (NO)}}\\
			\hline
			$\mathbf{M_D}$ & $\mathbf{M_{RS}}$ & $\mathbf{M_S}$ & $\mathbf{m_\nu}$(eV) & $\mathbf{m_{N}}$ & $\mathbf{m_{S}}$ \\
			\hline
			$\mathbf{\{10.00}$, ~~~~~~~ $\mathbf{36.37}$, ~~~~~~~ $\mathbf{81.33\}}$  & $\mathbf{\{2.24\times 10^{14}}$, $\mathbf{8.13\times 10^{14}}$, $\mathbf{3.59\times 10^{15}\}}$  & $\mathbf{\{5.00\times 10^{15}}$, $\mathbf{6.61\times 10^{15}}$, $\mathbf{2.57\times 10^{16}\}}$ & $\mathbf{\{0.010}$, $\mathbf{0.013}$, $\mathbf{0.051\}}$  & $\mathbf{\{1\times10^{13}}$, $\mathbf{1\times10^{14}}$, $\mathbf{5\times 10^{14}\}}$ & $\mathbf{\{5.00\times 10^{15}}$, $\mathbf{6.61\times 10^{15}}$, $\mathbf{2.57\times 10^{16}\}}$ \\
			\hline
			$\{0.32$, ~~~~~~~ ${1.15}$, ~~~~~~~ ${5.07\}}$  & ${\{2.24\times 10^{11}}$, ${8.13\times 10^{11}}$, ${3.59\times 10^{12}\}}$  & ${\{5.00\times 10^{12}}$, ${6.61\times 10^{12}}$, ${2.57\times 10^{13}\}}$ & $~~$ $\{0.010$,  $~~~$ $~~~~0.013$,$~~~$ $0.051\}$  & ${\{1\times10^{10}}$, ${1\times10^{11}}$, ${5\times 10^{11}\}}$ & ${\{5.00\times10^{12}}$, ${6.61\times10^{12}}$, ${2.57\times 10^{13}\}}$ \\
			\hline
			$\{1.00\times10^{-2}$, $3.64 \times10^{-2}$, $1.60\times10^{-1}\}$   & $\{2.24\times10^8$, $8.13\times10^8$, $3.59\times10^9\}$   & $\{5.00\times10^9$, $6.61\times 10^{9}$, $2.57\times 10^{10}\}$ & $~~$ $\{0.010$,  $~~~$ $~~~~0.013$,$~~~$ $0.051\}$  & $\{1\times10^7$, $1\times 10^{8}$, $5\times 10^{8}\}$ & $\{5.00\times10^9$, $6.61\times 10^{9}$, $2.57\times 10^{10}\}$ \\
			\hline \hline
			\multicolumn{6}{|c|}{\textbf{Inverted Ordering (IO)}}\\
			\hline
			$\mathbf{M_D}$ & $\mathbf{M_{RS}}$ & $\mathbf{M_S}$ & $\mathbf{m_\nu}$(eV) & $\mathbf{m_{N}}$ & $\mathbf{m_{S}}$ \\
			\hline
			$\mathbf{\{22.47}$, ~~~~~~~ $\mathbf{71.58}$, ~~~~~~~ $\mathbf{70.71\}}$  & $\mathbf{\{2.24\times 10^{14}}$, $\mathbf{7.12\times 10^{14}}$, $\mathbf{7.04\times 10^{14}\}}$  & $\mathbf{\{5.00\times 10^{15}}$, $\mathbf{5.07\times 10^{15}}$, $\mathbf{9.90\times 10^{14}\}}$ & $\mathbf{\{0.050}$, $\mathbf{0.051}$, $\mathbf{0.010\}}$  & $\mathbf{\{1\times10^{13}}$, $\mathbf{1\times10^{14}}$, $\mathbf{5\times 10^{14}\}}$ & $\mathbf{\{5.00\times 10^{15}}$, $\mathbf{5.07\times 10^{15}}$, $\mathbf{9.90\times 10^{14}\}}$ \\
			\hline
			$\{0.71$, ~~~~~~~ ${2.26}$, ~~~~~~~ ${2.24\}}$  & ${\{2.24\times 10^{11}}$, ${7.12\times 10^{11}}$, ${7.04\times 10^{11}\}}$  & ${\{5.00\times 10^{12}}$, ${5.07\times 10^{12}}$, ${9.90\times 10^{11}\}}$ & ${\{0.050}$, ${0.051}$, ${0.010\}}$  & ${\{1\times10^{10}}$, ${1\times10^{11}}$, ${5\times 10^{11}\}}$ & ${\{5.00\times10^{12}}$, ${5.07\times10^{12}}$, ${9.90\times 10^{11}\}}$ \\
			\hline
			$\{2.25\times10^{-2}$, $7.16 \times10^{-2}$, $7.07\times10^{-2}\}$   & $\{2.24\times10^8$, $7.12\times10^8$, $7.04\times10^8\}$   & $\{5.00\times10^9$, $5.07\times 10^{9}$, $9.90\times 10^{8}\}$ & $\{0.050$, $0.051$, $0.010\}$  & $\{1\times10^7$, $1\times 10^{8}$, $5\times 10^{8}\}$ & $\{5.00\times10^9$, $5.07\times 10^{9}$, $9.90\times 10^{8}\}$ \\
			\hline
		\end{tabular}
		\caption{A representative set of model parameters in Left-Right Symmetric Models and the order of magnitude estimation of various neutrino masses within the double seesaw mechanism. All the masses are expressed in units of GeV except the light neutrino masses, which are in the eV scale. Note: The first row for each mass ordering in bold format are the benchmark point values being used in numerical calculations. Here, the values of $m_{N_i}$ and $m_{1}$ in NO ($m_3$ in IO) are fixed as per the requirement of thermal leptogenesis, and the remaining parameters are obtained accordingly.}
		\label{tab:tabbp}
	\end{table}

	\subsubsection{Inverted Ordering}
	\label{subsec:IO1}
\begin{figure}[!h]
		\centering
		\hspace{-0.2 in}\includegraphics[width=0.98\textwidth]{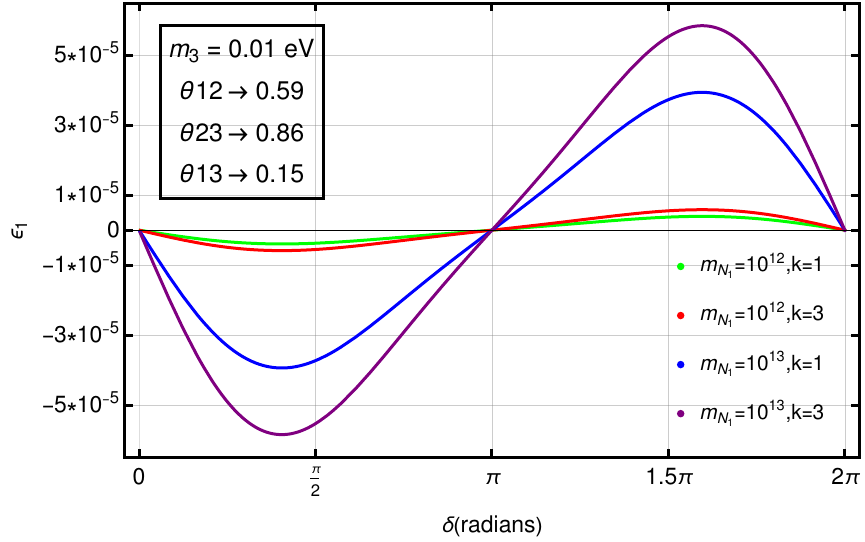}
		\caption{\textbf{(IO case)} Plot for the dependence of $\epsilon_1$ on CP-violating Dirac phase,~$\delta$ for different combinations of right-handed neutrino mass and the hierarchy in that sector. Value of variable $k$ represents the masses of heavier right-handed neutrino for the structure: $M_{N_2}=1\times10^{k}\times M_{N_1}$ and $M_{N_3}=5\times10^{k}\times M_{N_1}$.
		}
		\label{fig:3I}
	\end{figure}

	Similar to the analysis performed in subsection \ref{subsec:NO1}, we here present the results for the case of inverted ordering. Thus, from Eq.~(\ref{eq:cp1}), we get:
	\begin{equation}
		\begin{split}
			& \text{Im}[(M_D^\dagger M_D)^2_{21}]=  \sin{\delta}\left(9.46+2.36\cos{\delta}\right)\times10^{5}\text{~GeV}^2\\
			& \text{Im}[(M_D^\dagger M_D)^2_{31}]=  \sin{\delta}(-9.46+1.75\cos{\delta})\times10^{5}\text{~GeV}^2\\
			& (M_D^\dagger M_D)_{11}=  2003.03 \text{~GeV}^2
			\label{eq:analasymmI}
		\end{split}
	\end{equation}
	Here, we have used the benchmark point values $m_{N_1}=10^{13}\text{ GeV}$, $m_{N_2}=1\times10^{14}\text{ GeV}$, $m_{N_3}=5\times10^{14}\text{ GeV}$ and $m_{3}=0.01\text{ eV}$ of input parameters from the Table~\ref{tab:tabbp}~(Row 1 in IO) for the calculations. In Fig.~\ref{fig:3I}, we plot the variation of $\epsilon_1$ against the allowed range of CP-violating phase $\delta~[0,2\pi]$. From the plot, the dependence of $\epsilon_1$ on this Dirac CP phase is evident again, and the sinusoidal nature of this dependence,~(as expressed in Eq.~(\ref{eq:analasymmI})) is visible from the vanishing asymmetry value at angles 0, $\pi$ and $2\pi$. Also, the sign of the obtained $\epsilon_1$ value alters after $\delta=\pi$ radians and this again sets a limit on the allowed parameter space from the final baryon asymmetry requirements.	
	For a fair comparison, keeping the values of benchmark  input parameters the same as that for the NO case along with the best-fit value of $\delta=1.58\pi$~\cite{deSalas:2020pgw}, the asymmetry parameter for the IO case is numerically obtained as:
	\begin{equation}
		\epsilon_1 \sim +3.92\times 10^{-5}.
		\label{eq:asymIO}
	\end{equation}
	Comparing it with the obtained value of $\epsilon_1$ in Eq.~(\ref{eq:asymNO}) for the NO case, we see a sign and magnitude change in its value. Also, by using all the relevant parameters mentioned above, the generic structure of $M_D$ as given in Eq.~(\ref{MDMatrix}) is obtained numerically for the IO case as,
	\begin{equation}
		M_D= \begin{pmatrix}
			-0.76 + 35.45i & ~-11.76 + 15.57i & ~-10.12 - 16.22i\\
			-11.76 + 15.57i & ~-3.02 + 62.45i & ~~~~0.69 + 6.07i\\
			-10.12 - 16.22i & ~~~0.69 + 6.07i & ~~~~~3.42 + 62.33i
		\end{pmatrix}
		\label{eq:cpmatI}
	\end{equation}
\begin{figure}
	\centering
	\begin{subfigure}[b]{0.73\textwidth}
		\includegraphics[width=1\linewidth]{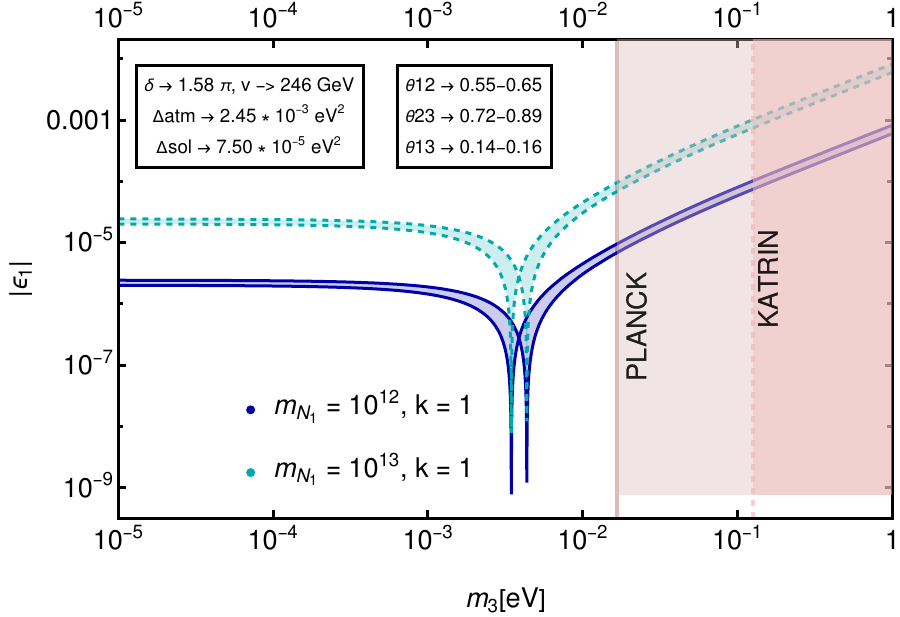}
		\caption{}
		\label{fig:Ng1I}
	\end{subfigure}
	
	\begin{subfigure}[b]{0.73\textwidth}
		\includegraphics[width=1\linewidth]{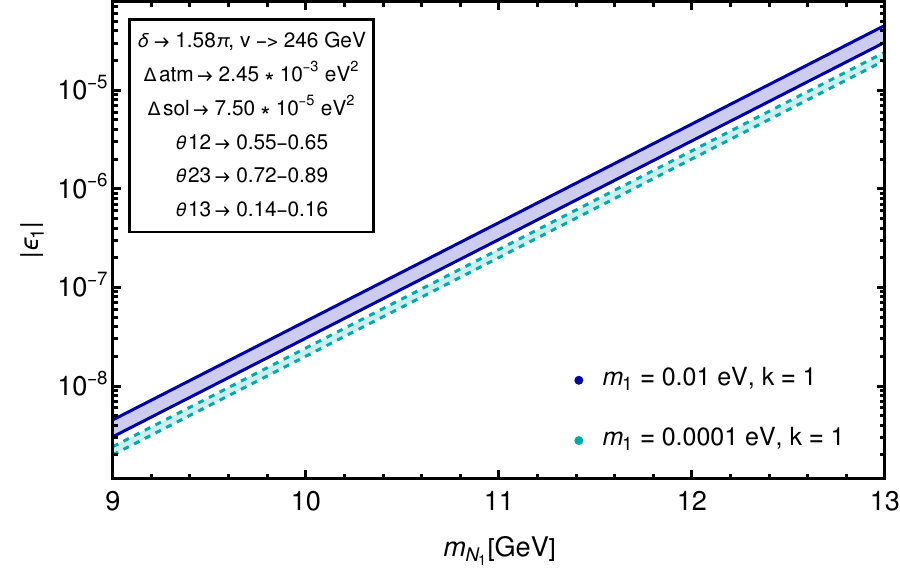}
		\caption{}
		\label{fig:Ng2I}
	\end{subfigure}
	
	\caption[]{\textbf{(IO case)} The above figure shows two region plots depicting the dependence of asymmetry parameter magnitude,~$|\epsilon_1|$ on lightest active neutrino mass,~$m_3$ in plot~(a) and on lightest right-handed neutrino mass,~$m_{N_1}$ in plot~(b). For both the plots we vary the values of oscillation parameters $\theta_{12}$, $\theta_{23}$ and $\theta_{13}$ within their allowed $3\sigma$ range. CP phase $\delta$ has been set to its best-fit value of $1.08\pi$. Value of variable $k$ represents here the masses of heavier right-handed neutrino for the structure: $M_{N_2}=1\times10^{k}\times M_{N_1}$ and $M_{N_3}=5\times10^{k}\times M_{N_1}$. The vertical pink bands in plot~(a) represent the bound corresponding to the upper limit on the sum of light neutrino masses of $0.12~\text{eV}$ reported by the Planck~\cite{Planck:2018vyg} and the prospective bound of $0.20~\text{eV}$ that can be set by the KATRIN~\cite{KATRIN:2019yun} collaboration.}
	\label{fig:4I}
\end{figure}
\noindent We now plot the dependence of the CP asymmetry parameter ($|\epsilon_1|$) on the lightest active neutrino mass, $m_{3}$ and on the lightest right-handed neutrino mass, $m_{N_1}$, respectively in figures \ref{fig:Ng1I} and~\ref{fig:Ng2I}, for the IO case. For both the plots, we vary the values of oscillation parameters $\theta_{12}$, $\theta_{23}$ and $\theta_{13}$ in a similar manner as done in NO case. In Fig.~\ref{fig:Ng1I}, we see that the value of $\epsilon_1$ changes sign at some singularity point within $10^{-3}-10^{-2}\text{ eV}$ range of $m_3$ for both the plot points of $m_{N_1}=10^{12}\text{ GeV}$~(denoted by blue band region) and $m_{N_1}=10^{13}\text{ GeV}$~(denoted by cyan band region). We find that for the considered parameter space, the value of $\epsilon_1$ is obtained to be negative before the singularity point and positive after the singularity point for both of the $m_{N_1}$ values. In Fig.~\ref{fig:Ng2I}, we see that the magnitude of $\epsilon_1$ directly depends on the value of $m_{N_1}$ for the entire range of interest but in contrast to the behaviour of $\epsilon_1$ in Fig.~\ref{fig:Ng2}, the sign of $\epsilon_1$ is governed by the choice of $m_1$. For the case, $m_1=0.01\text{ eV}, k=1$~(the blue band region), the value of obtained $\epsilon_1$ has a positive sign and for the case, $m_1=0.0001\text{ eV}, k=1$~(the cyan band region), the value of obtained $\epsilon_1$ has a negative sign. The values of other relevant input parameters are given within the plots, and the CP-violating phase is set to its best-fit value of $\delta=1.58\pi$ for the IO case.
	
\begin{figure*}[]
		\includegraphics[width=0.46\textwidth]{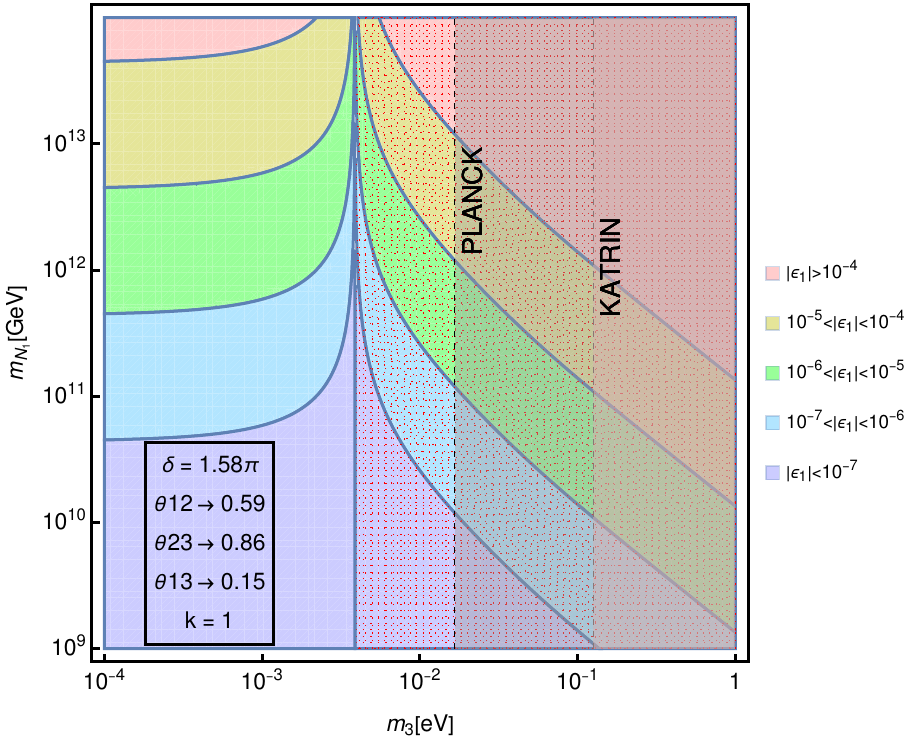}
		\includegraphics[width=0.54\textwidth]{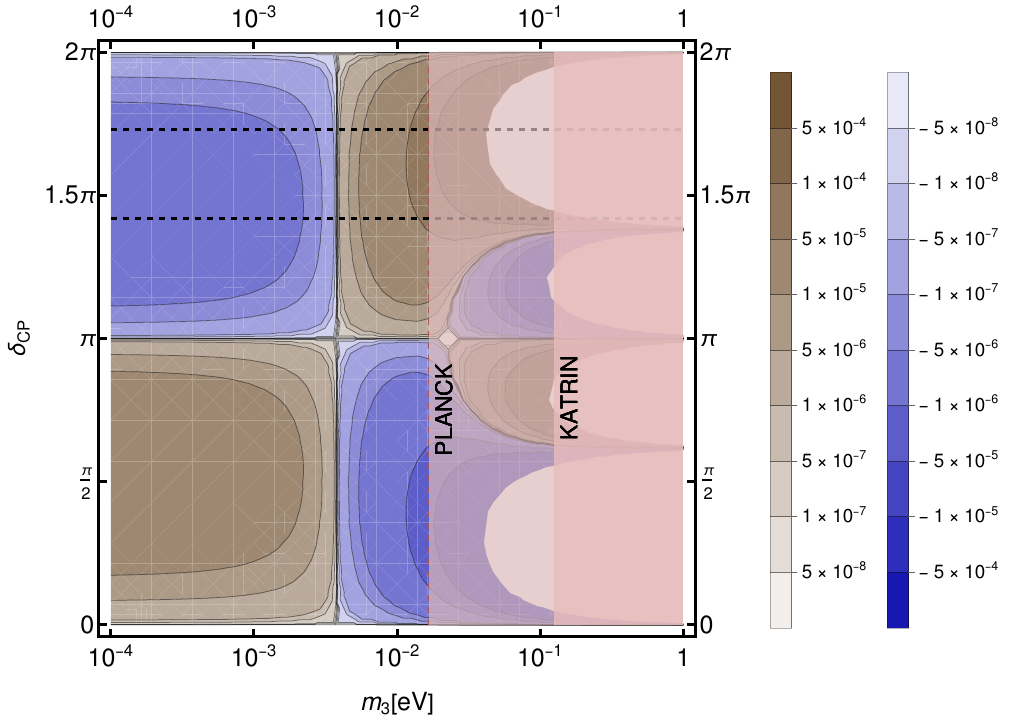}
		\caption{\textbf{(IO case)} The region plots here explore the dependence of asymmetry parameter $(\epsilon_1)$ on a set of other input parameters. In the left plot, input variables are the masses of lightest active neutrino and lightest right-handed neutrinos, respectively on the x and y axis. The values of other relevant parameters are mentioned within the plot. The dotted red mesh region in the plot shows the parameter space where the value of $\epsilon_1$ has a positive sign. In the right plot, input variables are the mass of lightest active neutrino and the value of Dirac CP phase, respectively on x and y axis. All other relevant parameters are fixed at their best-fit values given in Table~\ref{tab:exp}. The masses of right-handed neutrinos are set at their benchmark values of $m_{N_1}=1\times10^{13}~\text{GeV}$, $m_{N_2}=1\times10^{14}~\text{GeV}$ and $m_{N_3}=5\times10^{14}~\text{GeV}$. For both the plots, we also show the experimental bounds on lightest active neutrino mass from Planck and KATRIN experiments marked by vertical bands. Also, in the right plot, the black dashed lines running horizontally marks the currently accepted $1\sigma$ band for Dirac CP phase ($\delta_{CP}$)~\cite{deSalas:2020pgw}.
		}
		\label{fig:9I}
	\end{figure*}

	In Fig.~\ref{fig:9I}, we show two region plots depicting the dependence of $\epsilon_1$ in a $m_3-m_{N_1}$ plane and in a $m_3-\delta_{\text{CP}}$ plane, respectively in the left and the right plot. Bands of different $\epsilon_1$ values are obtained for the combination of input parameter values in both the plots. In the left plot, we see that for a particular value of $m_3$ between $10^{-3}\text{ eV}$ to $10^{-2}\text{ eV}$, the value of $\epsilon_1$ changes sign and this point of singularity remains unaffected from the value of $m_{N_1}$ for the entire mass range. The plot region covered in dotted red mess is the region where the value of $\epsilon_1$ is obtained to be positive. In the right plot, we obtain various contours for different values of $\epsilon_1$ ranging from $-5.0\times10^{-4}$ to $+5.0\times10^{-4}$ for the given plot range. The features of these contours are a direct consequence of the combined effect of: 1) the sinosoidal dependence of $\epsilon_1$ on the $\delta$ parameter and, 2) the sign flipping dependence of $\epsilon_1$ on the lightest active neutrino mass,~$m_3$. The blue contours in the plot represents the set of $\delta$ and $m_3$ values for which we obtain a negative $\epsilon_1$ value and a positive value is obtained for the brown contours.
		
	\newpage
	
\begin{figure}[!htb]
		\centering
		\begin{subfigure}[b]{0.90\textwidth}
			\includegraphics[width=1\linewidth]{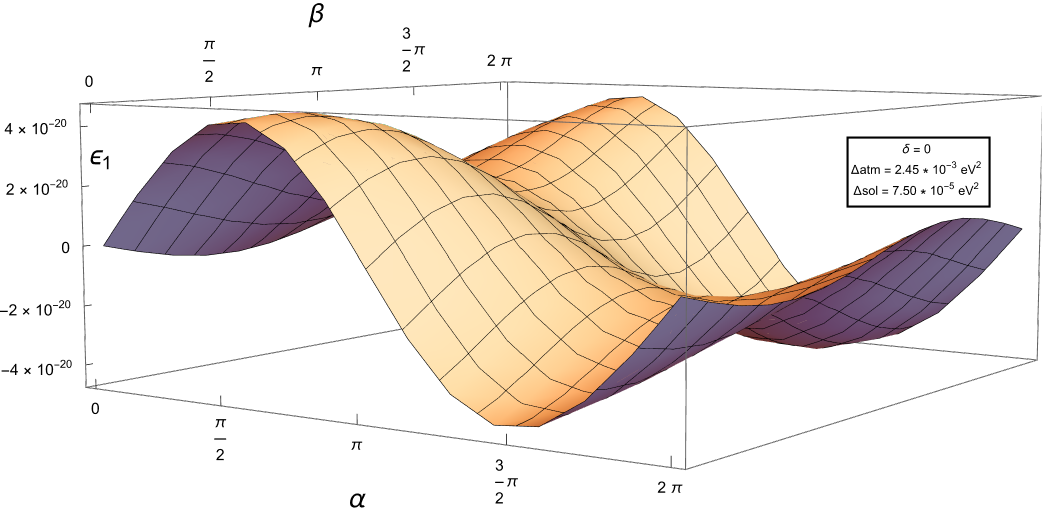}
			\caption{}
			\label{fig:IOMPA} 
		\end{subfigure}
		
		\begin{subfigure}[b]{0.85\textwidth}
			\includegraphics[width=1\linewidth]{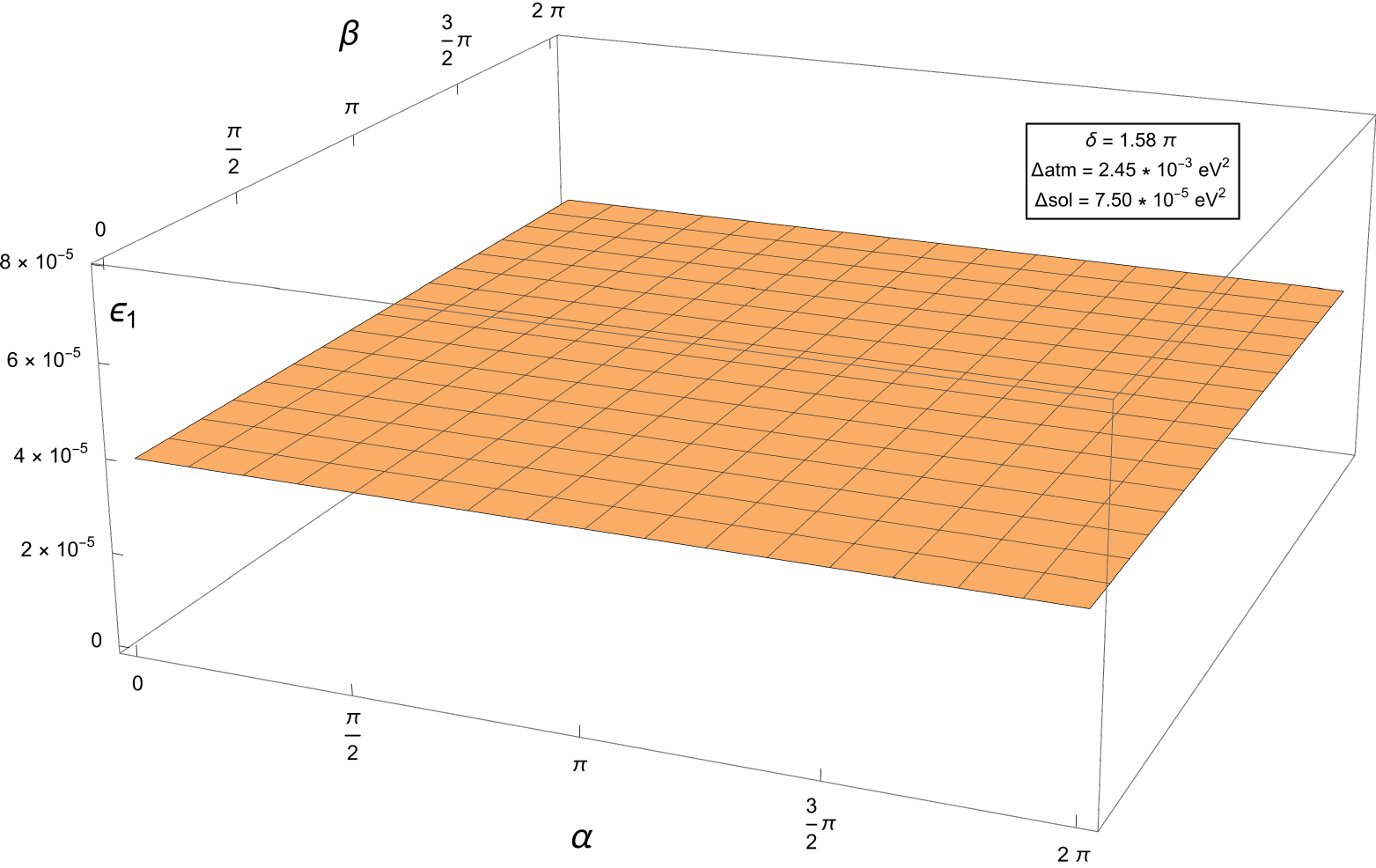}
			\caption{}
			\label{fig:IOMPB}
		\end{subfigure}
		
		\caption{\textbf{(IO case)} The above figure shows two 3D plots (a) and (b) depicting the dependence of asymmetry parameter~$\epsilon_1$ on both the Majorana phases $\alpha$ and $\beta$ for the case of zero~$(\delta=0)$ and best-fit~$(\delta=1.58\pi)$ Dirac phase, respectively. The values of $\alpha$ and $\beta$ run from $0$ to $2\pi$ for both the plots. In plot (a), the Dirac CP phase is set at zero and thus we obtain a $(\alpha, \beta)$ dependent region of $\epsilon_1$ values~(light copper colored region). In plot (b), we see that there is almost no dependence of $\epsilon_1$ on the Majorana phases and thus a flat plane~(orange colored region) is obtained for the entire plot range. The values of all the other relevant input parameters are mentioned within these plots. The range of obtained $\epsilon_1$ values along with its sign are shown on the z-axis in these plots.}
		\label{fig:IOMP}
	\end{figure}

	\underline{\textbf{Inclusion of Majorana Phases}}\\[.1in]
	Similar to the analysis done in subsection \ref{subsec:NO1}, we here incorporate the Majorana phases ($\alpha, \beta$) and express the various terms in Eq. (\ref{eq:cp1}) analytically for two different values of the Dirac phase: \textbf{(1.)} $\delta=0$ (CP conserving) and \textbf{(2.)} $\delta=1.58\pi$ (best-fit).
	\begin{enumerate}
		\item[ \textbf{(1.)}] For $\delta=0$.
		\begin{equation}
			\begin{split}
				& \text{Im}[(M_D^\dagger M_D)^2_{21}]\simeq  -(5.87\sin{\alpha}+0.98\sin{\beta}+3.91\sin{(\alpha-\beta)})\times10^{-10}\text{~GeV}^2\\&\hspace{1.19in}+f(\alpha,\beta)\,\mathcal{O}(10^{-27})\\
				& \text{Im}[(M_D^\dagger M_D)^2_{31}]\simeq  -(1.48\sin{\alpha}+0.74\sin{\beta}-5.93\sin{(\alpha-\beta)})\times10^{-10}\text{~GeV}^2\\&\hspace{1.19in}+f(\alpha,\beta)\,\mathcal{O}(10^{-26})\\
				&(M_D^\dagger M_D)_{11}= [2003.03-[ 1.71\cos{\alpha}-1.14\cos{(\alpha-\beta)}]\times10^{-13}]\text{~GeV}^2  
			\end{split}
		\end{equation}
		With the values of input benchmark parameters, the asymmetry parameter here is numerically obtained as:
		\begin{equation}
			\epsilon_1 \sim [(3.03+0.29 \cos{\alpha})\sin{\alpha}+1.34 \sin{(\alpha-\beta)}+0.55 \sin{\beta}]\times 10^{-20}.
		\end{equation}
		
		\item[ \textbf{(2.)}] For $\delta=1.58\pi$.
		\begin{equation}
			\begin{split}
				& \text{Im}[(M_D^\dagger M_D)^2_{21}]\simeq  -9.73\times10^5\text{~GeV}^2 +f(\alpha,\beta)\,\mathcal{O}(10^{-10})\\
				& \text{Im}[(M_D^\dagger M_D)^2_{31}]\simeq  8.74\times10^5\text{~GeV}^2+f(\alpha,\beta)\,\mathcal{O}(10^{-10})\\
				& (M_D^\dagger M_D)_{11}\simeq 2003.03\text{~GeV}^2+f(\alpha,\beta)\,\mathcal{O}(10^{-13})  
			\end{split}
		\end{equation}
		With the values of the input benchmark parameters, the asymmetry parameter is numerically obtained as follows: 
		\begin{equation}
			\epsilon_1 \sim +3.92\times 10^{-5}+f(\alpha,\beta)\,\mathcal{O}(10^{-20}).
		\end{equation}
	\end{enumerate}
	Keeping our analysis streamlined with the NO case explored in subsection \ref{subsec:NO1}, we again investigate the variation of $\epsilon_1$ on Majorana phases for IO case in Fig \ref{fig:IOMP}. From the figures \ref{fig:IOMPA} and \ref{fig:IOMPB}, one may see no qualitative dependence in the behaviour of $\epsilon_1$ on the Majorana phases for both $\delta=0$ (CP conserving) and $\delta=1.58\pi$ (best-fit) cases.   
	
	Based on our investigations conducted here and in subsection \ref{subsec:NO1} concerning the inclusion of Majorana phases for IO and NO mass spectra of active neutrinos, respectively, we have determined that, within our model framework, the contribution of Majorana phases to CP asymmetry is negligible compared to the Dirac CP phase. This observation also implies that the scale of leptogenesis in our framework for both NO and IO cases remains unaffected by the inclusion of Majorana phases. Consequently, in the subsequent discussions, we confine our analysis to the Dirac CP phase ($\delta$) as the sole source of CP violation, i.e. setting Majorana phases ($\alpha,\beta$) equal to zero.

	\subsection{Cosmological Evolution of asymmetry}
	\label{subsec:CEA}
	In this subsection, we discuss the various processes that thread together to create the observed baryon asymmetry of the Universe via leptogenesis. The right-handed neutrinos $(N_R)$ are present at the high-energy scales and decay to their light counterparts as the temperature of the Universe cools down. It is possible for all the three generation of RH neutrinos to decay in a CP-violating way, thus producing an asymmetry in the leptonic sector. Given a mass hierarchy between them, it is natural for the $N_1$s to decay at the very last. The decay width of such a process is expressed in Eq.~(\ref{GammaTotal}). The asymmetries produced via the decay of $N_3$ and $N_2$ wash off entirely as the Universe's temperature cools below the mass scale of $N_2$.
	
	Particularly for our framework, there is an additional contribution to the number density of $N_R$ coming from the decay of $S_L$ neutrinos from the second Yukawa term in Eq.~(\ref{eq:lg1}). Although as the mass scale of $S_L$ is higher than the mass of $N_1$ so, these interactions do not play a significant role in producing the lepton asymmetry but only alter the abundance of $N_1$. This leaves the decay of $N_1$ quite significant for the process of leptogenesis. The first Yukawa term in Eq.~(\ref{eq:lg1}) with complex $Y_D$ values provides the necessary net asymmetry in the production of active neutrinos and leptons.
	
	The initial number density of RH neutrinos, the contribution of processes that lead to a washout of created asymmetry, and various other scatterings of $N_1$ all play an essential role in deciding the final cosmological abundance of $N_1$ and other leptons. Mathematically, such a dynamic interplay is dealt with by solving the coupled Boltzmann equations (BEs) for all these particles. The SM lepton asymmetry is transferred into baryon asymmetry via the well established electroweak sphalerons. The symmetric baryon component is almost entirely wiped out by hadronic annihilations, and only the asymmetric component survives.
	
	The final baryon asymmetry can be accounted from the result of a competition between the production and washout processes. These processes in themselves usually encompass the decay and inverse decay of $N_1$ and the off-shell $\Delta L=0$ and $\Delta L=2$ scatterings~\cite{Luty:1992un,Plumacher:1996kc}, given no degeneracy in the mass hierarchy of heavy neutrinos, like in the case of resonant leptogenesis. Now we present the most general structure of coupled BEs required for our analysis, neglecting the contributions of $S_L$, $N_2$ and $N_3$ decays for the abovementioned reasons. These equations are flavor singular and are in direct comparison with the results of~\cite{Frere:2008ct,Deppisch:2010fr}.
	
	\begin{align}
		\frac{d\eta^{N_1}}{dz} \ & = \ -\left(\frac{\eta^{N_1}}{\eta^N_{\rm eq}}-1\right)(D_1+S_1) \;, \label{be1} \\
		\frac{d\eta^{\Delta L}}{dz} \ & = \varepsilon_{1}\left(\frac{\eta^{N_1}}{\eta^N_{\rm eq}}-1\right)\widetilde{D}_1 -\frac{2}{3} \eta^{\Delta L} W_l \;, \label{be2}
	\end{align}
	where 
	$z=m_{N_1}/T$ is a dimensionless variable ($T$ being the temperature of the Universe) and
	$\eta^N_{\rm eq}  \equiv  n^N_{\rm eq}/n^\gamma  =  z^2 K_2(z)/2\zeta(3) $
	is the heavy neutrino equilibrium number density, $K_n(z)$ being the $n$-th order modified Bessel function of the second kind and $\zeta(3)$ is Riemann zeta function ($\zeta(s)$) evaluated at $s=3$. $n^\gamma$ is referred as the comoving photon number density. The various decay ($D_1,~\widetilde{D}_1$), scattering ($S_1$) and washout ($W_l$) rates appearing in equations (\ref{be1}) and~(\ref{be2}) are given by
	
	\begin{align}
		\widetilde{D}_1 & = \frac{z}{n^\gamma H_N} \widetilde{\gamma}^D,~~~~~ D_1   =  \frac{z}{n^\gamma H_N} \gamma^D,~~~~~ S_1  =  \frac{z}{n^\gamma H_N} (\gamma^{S_L} + \gamma^{S_R}), \label{Dtilde}\\[0.1in]
		W_l & = \frac{z}{n^\gamma H_N} \left[ \gamma^D + \widetilde{\gamma}^{S_L}+ \widetilde{\gamma}^{S_R} + \gamma^{(\Delta L = 0)} + \gamma^{(\Delta L = 1)} + \gamma^{(\Delta L = 2)} \right], \label{wash}
	\end{align}
	where $H_N\equiv H(z=1)\simeq 17 m_{N_1}^2/M_{\text {Pl}}$ is the Hubble parameter at $z=1$, assuming only SM degrees of freedom in the thermal bath, $M_{\text {Pl}}=1.2\times 10^{19}$ GeV is the Planck mass.
	The definitions of various decay rates $(\gamma)$ that are involved here are given explicitly in Appendix~\ref{app:ir}. Based on the analysis done in~\cite{Zhang:2020lir}, we may safely neglect the contributions of $\Delta L=2$ scattering processes of $N_1$ in the BEs. This result is derived in details in Ref. \cite{Giudice:2003jh}.

	\begin{figure*}[!ht]
		\centering
		\includegraphics[width=0.7\textwidth]{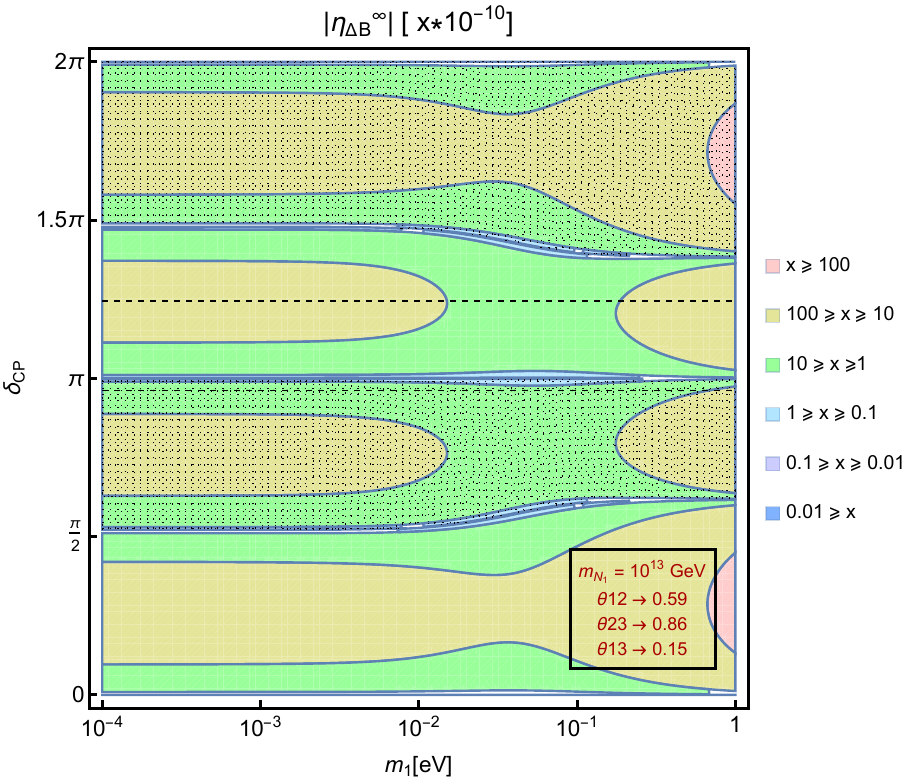}
		\caption{\textbf{(NO case)} Scatter plot for the obtained value of final baryon asymmetry ($\eta_{\Delta B}^{\infty}$) in the case of strong washout regime for normal ordering of active neutrinos. The right-handed neutrino hierarchy for the plot is set as: $M_{N_1}=1\times10^{13}\text{ GeV}$, $M_{N_2}=1\times10^{14}\text{ GeV}$ and $M_{N_3}=5\times10^{14}\text{ GeV}$. Plot region covered in black mesh shows the parameter space where the value of~$\eta_{\Delta B}^{\infty}$ has a negative sign. The black dashed lines running horizontally across the plot shows the currently accepted $1\sigma$ band for CP-violating parameter, $\delta$ for NO case. Values of other relevant parameters are mentioned within the plot.}
		\label{fig:11}
	\end{figure*}
	
	\begin{figure*}[!ht]
		\centering
		\includegraphics[width=0.7\textwidth]{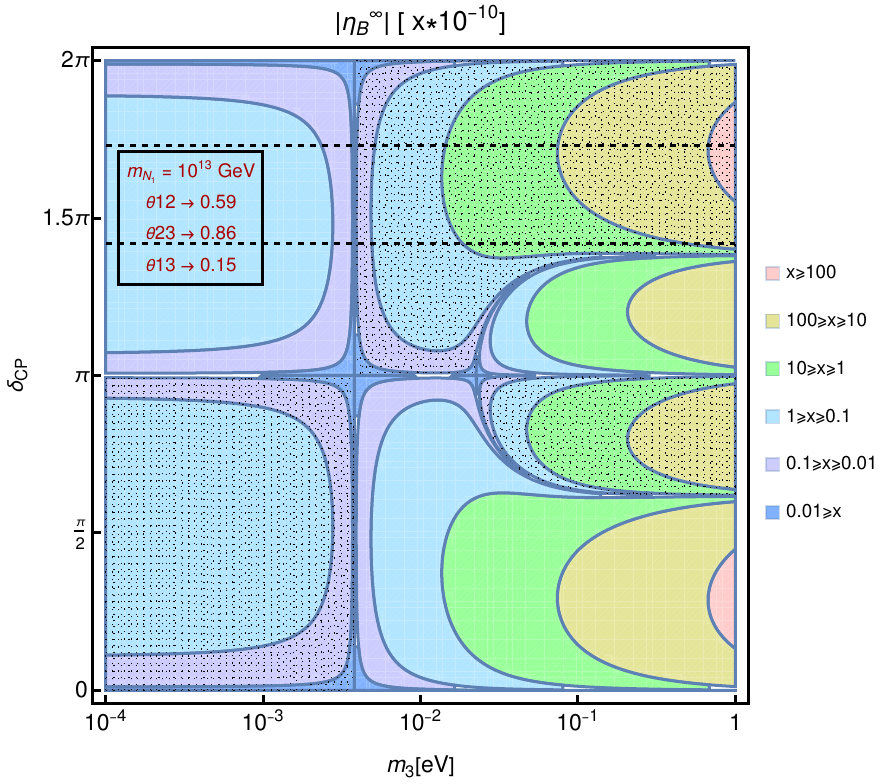}
		\caption{\textbf{(IO case)} Scatter plot for the obtained value of final baryon asymmetry ($\eta_{\Delta B}^{\infty}$) in the case of strong washout regime for inverted ordering of active neutrinos. The right-handed neutrino hierarchy for the plot is set as: $M_{N_1}=1\times10^{13}\text{ GeV}$, $M_{N_2}=1\times10^{14}\text{ GeV}$ and $M_{N_3}=5\times10^{14}\text{ GeV}$. Plot region covered in black mesh shows the parameter space where the value of~$\eta_{\Delta B}^{\infty}$ has a negative sign. The black dashed lines running horizontally across the plot shows the currently accepted $1\sigma$ band for CP-violating parameter, $\delta$ for IO case. Values of other relevant parameters are mentioned within the plot.}
		\label{fig:11I}
	\end{figure*}
	
	\subsection{Results and discussions}
	\label{subs:rad}
	In this subsection, we dive into the in-depth analysis of analytical and numerical aspects of asymmetry evolution. For a comprehensive analysis, we solve the two coupled BEs (\ref{be1}) and~(\ref{be2}), including the various decay, scattering, and washout terms mentioned in Appendix~\ref{app:ir}. The contribution of these terms decides the final fate of asymmetry in the leptonic sector. The most important input parameters that affect their relative strengths include $m_{N_1}$, $M_{W_R}$, and $Y_D$. From the references~\cite{Ma:1998sq,Frere:2008ct,Carlier:1999ac,Dhuria:2015cfa}, one implies that the analysis could be approached for two major cases depending on the relative mass scale of $N_1$ and $W_R$. The choice of lightest right-handed neutrino mass $(m_{N_1})$ has been set to $10^{13}\text{ GeV}$ in our benchmark analysis~(refer Table~\ref{tab:tabbp}) to set up a fair scale for thermal leptogenesis. Also, a decent mass hierarchy $m_{N_2}=10^{14}\text{ GeV}$ and $m_{N_3}=5\times10^{14}\text{ GeV}$ has been taken to ensure a safe neglect of $N_{2,3}$ decays to lepton asymmetry.
	
	For the case $M_{W_{R}}<m_{N_1}$, it can be seen from the analysis done in Ref.~\cite{Ma:1998sq} that to generate a successful asymmetry, a lower bound of $10^{16}\text{ GeV}$ is obtained on the mass of $N_1$ for the case of $g_L^2=g_R^2=0.4$. Such a high mass scale of $N_R$ pushes the mass scale of $S_L$ even further up due to the hierarchy of double-seesaw used in our framework. For the purpose of this work, we restrict ourselves to consider the following case.
	
	In our analysis, we take $M_{W_R}(>m_{N_1})=10^{15}\text{~GeV}$ along with the other benchmark point values. This choice of $M_{W_R}$ is carefully done to make the scattering processes of $W_R$'s into $e_R$'s via $N_R$ exchange depart away from thermal equilibrium. Taking inspiration from an analogous standard model process of $W_L$ scattering as given in Ref.~\cite{Sarkar:1996df}, the translated mathematical inequality for this statement is given as:
	\begin{equation}
		M_{W_R}\geq3\times10^{6}\text{~GeV}{\Big(\frac{m_{N}}{10^2\text{~GeV}}\Big)}^{\frac{2}{3}}
		\label{eq:wR}
	\end{equation}
	For $m_{N_1}=10^{13}\text{ GeV}$, this translates to $M_{W_R}\gtrsim6.55\times10^{13}\text{ GeV}$. This choice ensures that asymmetry created by $M_{N_1}$ is not washed out by these $W_R$ scattering processes. Based on the values in Table~\ref{tab:outofeq}, it is important to mention here that our framework dynamics leads us to work in a strong washout regime for any given set of input parameter values for both of the normal and inverted hierarchies of active neutrino masses. Thus, the following discussions in this subsection are relevant for both the hierarchies, until stated otherwise explicitly. In a strong washout regime, we assume that any initial asymmetry that we begin with, is washed out by the relatively outnumbered inverse decay processes~\cite{Davidson:2008bu}, and thus it is safe to take an initial zero asymmetry in the lepton sector. With the cosmic evolution, the temperature drops, and asymmetry begins to survive once the inverse decay processes of $N_1$ depart from thermal equilibrium. This could loosely be equated to a mathematical relation given as:
	\begin{equation}
		\Gamma(\phi L\rightarrow N_1)\simeq\frac{1}{2}\Gamma_De^{-m_{N_1}/T}<H
		\label{eq:iD}
	\end{equation}
	Here, $\Gamma_D$ is given from Eq.~(\ref{GammaTotal}) and $H$ is the Hubble's parameter. Using Eq.~(\ref{eq:iD}), it can be seen that below a specific temperature $T_f$, the $N_1$ density is Boltzmann suppressed and this remaining density can create the required lepton asymmetry despite strong washouts to begin with.
	
	\begin{figure}
		\centering
		
		\begin{subfigure}{0.48\textwidth}
			\includegraphics[width=\linewidth]{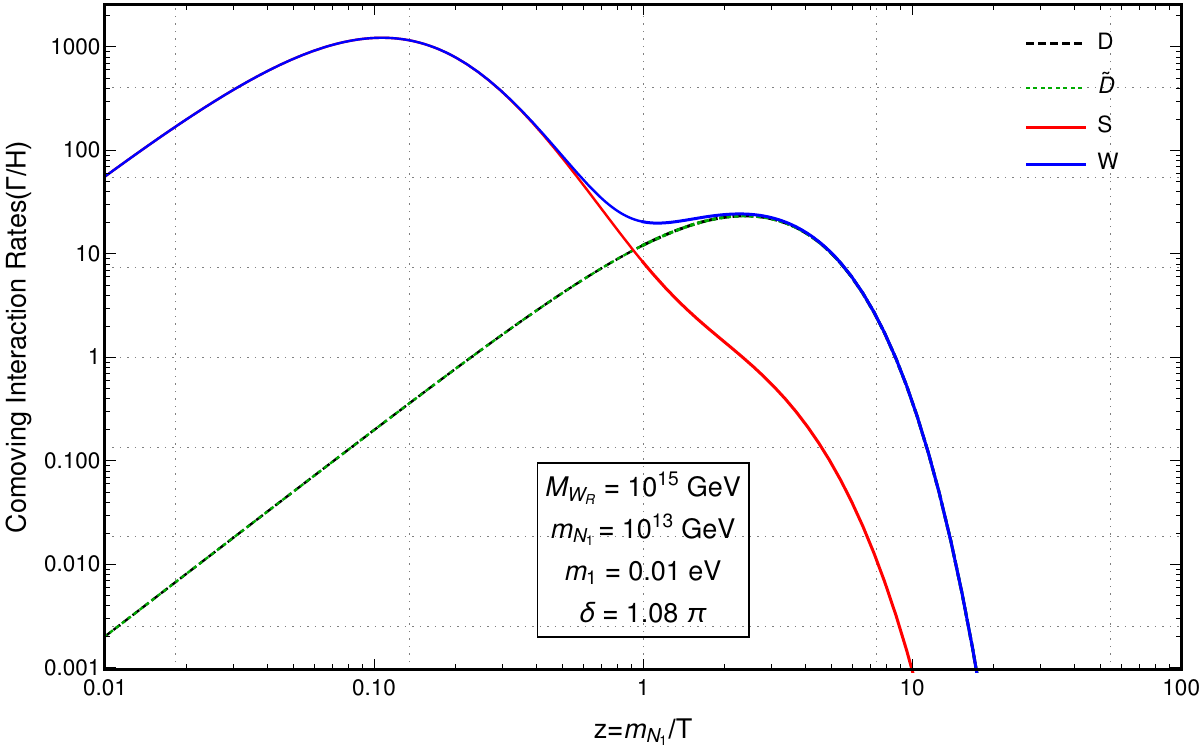}
			\caption{}
			\label{fig:10sub1}
		\end{subfigure}
		\hfill
		\begin{subfigure}{0.49\textwidth}
			\includegraphics[width=\linewidth]{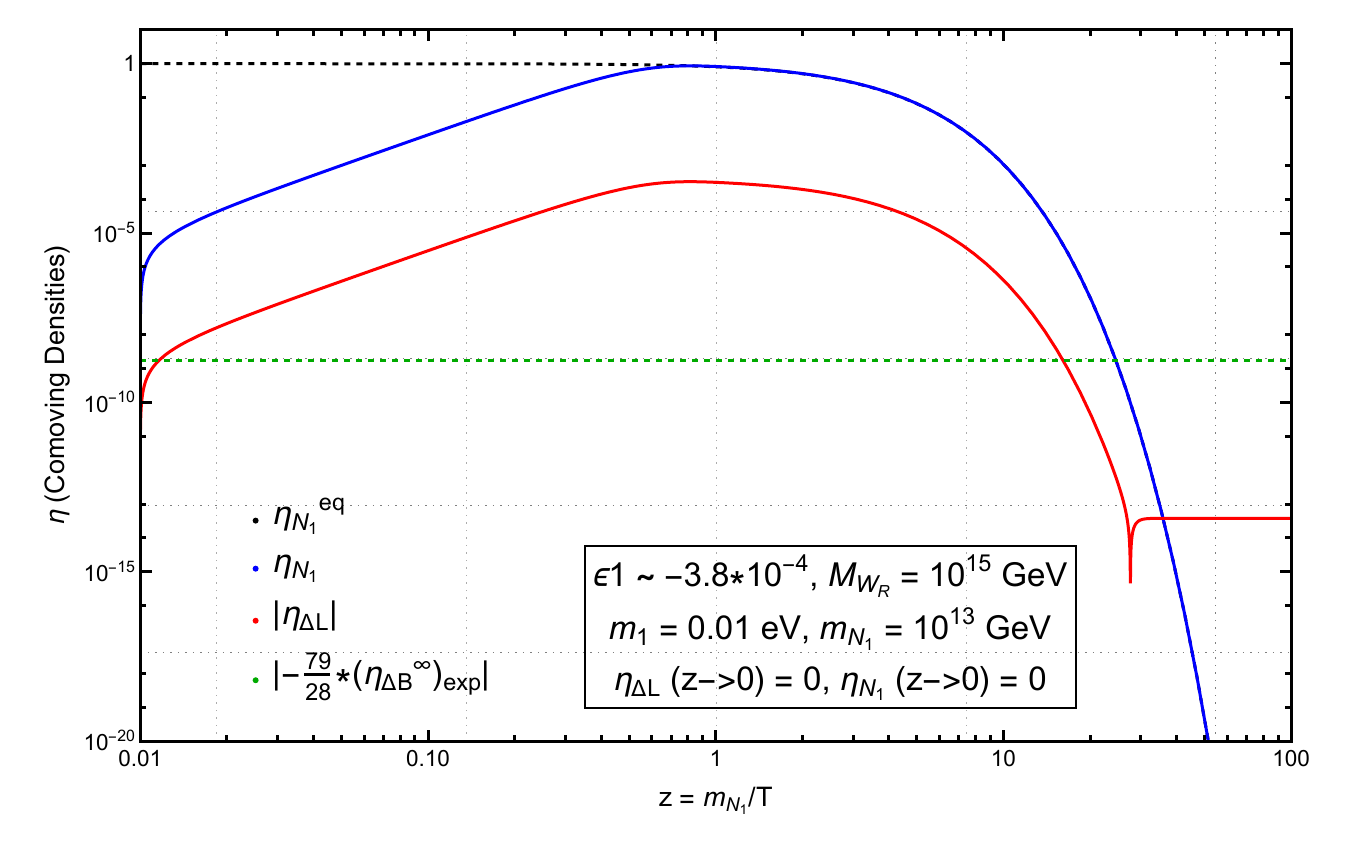}
			\caption{}
			\label{fig:10sub2}
		\end{subfigure}
		
		\medskip
		
		\begin{subfigure}{0.49\textwidth}
			\includegraphics[width=\linewidth]{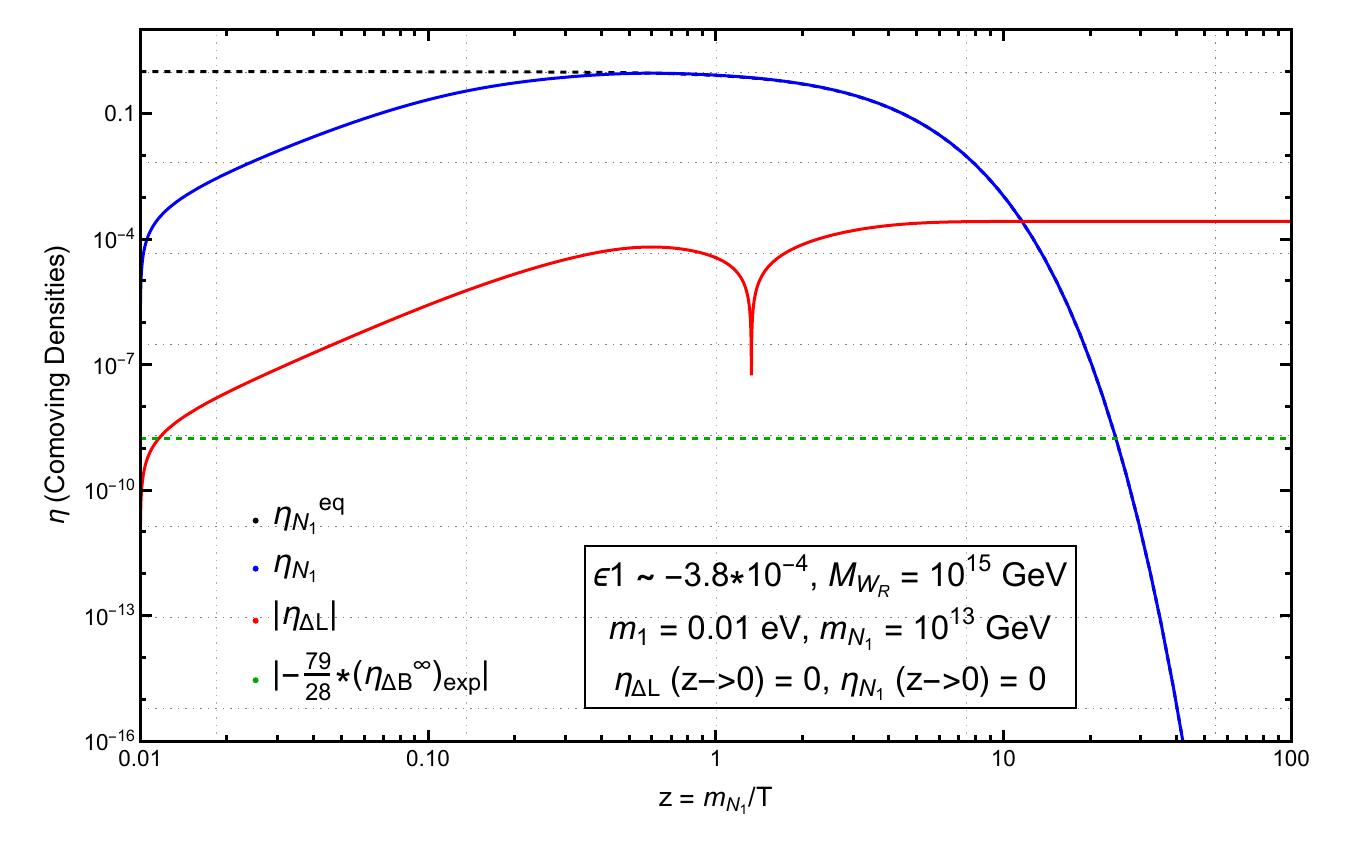}
			\caption{}
			\label{fig:10sub3}
		\end{subfigure}
		\hfill
		\begin{subfigure}{0.49\textwidth}
			\includegraphics[width=\linewidth]{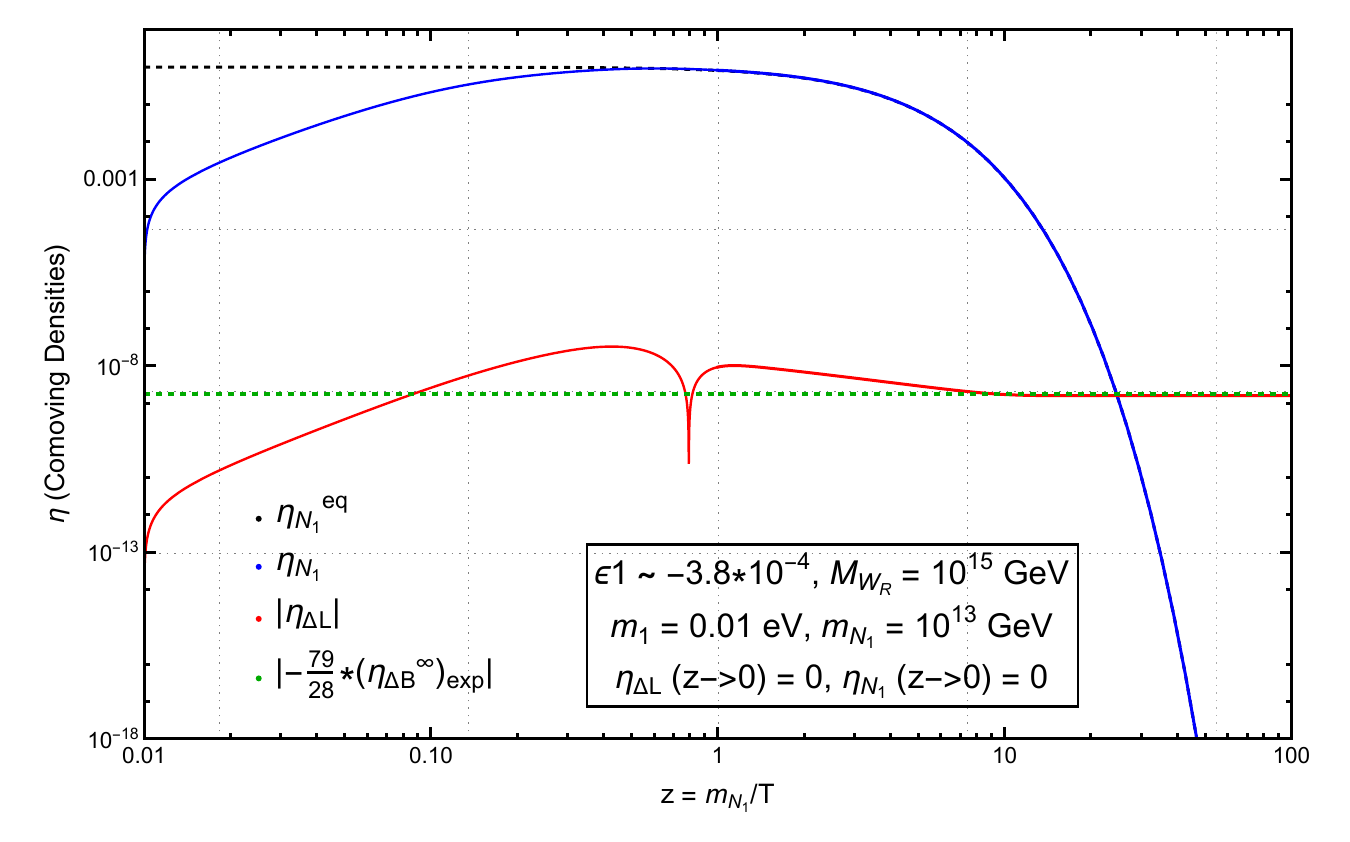}
			\caption{}
			\label{fig:10sub4}
		\end{subfigure}
		
		\caption{Figure shows four plots depicting the cosmological evolution of the various interaction terms and the number densities present in the coupled BEs (\ref{be1}) and (\ref{be2}) for the \textbf{NO} case. Plot (a) here shows the relative contributions of the different interaction terms that play a role in the evolution of lepton asymmetry. These terms include the $N_1$ decays~(black dashed line, denoted by D), inverse-decays~(green dotted line, denoted by $\tilde{D}$), scatterings involving leptons~(solid red line, denoted by S) and the asymmetry washout interactions~(solid blue line, denoted by W). Plot (b) shows the cosmological evolution of the lepton asymmetry number density with only decays and inverse-decays of $N_1$ considered. In Plot (c), we also include the effects of various scattering interactions involving leptons, while calculating the evolution of lepton asymmetry. Plot (d) shows the cosmological evolution of lepton asymmetry incorporating all the interactions present in the BEs i.e. decays, inverse-decays, scatterings and the terms that lead to asymmetry wash-out. The values of all the relevant input parameters and the considered initial conditions are mentioned within the plots. In the plots (b), (c) and (d), for a reference, we also show the cosmological evolution of the $N_1$ number density~($\eta_{N_1}$) represented by the solid blue line, the $N_1$ equilibrium number density~($\eta_{N_1}^{\text{eq}}$) represented by the dashed black line. In these 3 plots, the magnitude of the number density for the lepton asymmetry value~($|\eta_{\Delta\text{L}}|$) is denoted by the solid red line and the green dashed line represent the currently accepted asymmetry value by observations~\cite{Planck:2015fie}. All the oscillation parameters are fixed at their best-fit values.}
		\label{fig:10}
	\end{figure}

	The quantity $\Gamma_D/H$ is usually referred as the decay parameter and is denoted by $K$. Based on the value of $K$, an analytical approximation can be made for the final lepton asymmetry. In Ref.~\cite{Fong:2012buy}, such an approximation for the case of strong washout scenario is expressed as,
	\begin{equation}
		Y_{\Delta L}(\infty)=\frac{2}{z_fK}\epsilon_1Y_{\Delta L}^{eq}=\frac{\pi^2}{6z_fK}\epsilon_1Y_{N_1}^{eq}(0)
		\label{eq:Yanal}
	\end{equation}
	Here, $Y(=n/s)$ is the scaled number density of particles (with $s$ being the entropy density of the Universe) and is closely related to $\eta$ parameter, used in our analysis. Both quantities evolve similarly with temperature $(T)$ and thus can be used interchangeably with proper scaling ensured. Parameter $z_f$ usually lies between $7-10$ for $K=10-100$, and also it is assumed that initially, $N_1$ abundance is equal to its thermal abundance i.e., $Y_{N_1}\simeq Y_{N_1}^{eq}$. One may also find a more detailed analytical approximation of Eq.~(\ref{eq:Yanal}) in Ref.~\cite{Buchmuller:2004nz}. Using Ref.~\cite{Fong:2012buy}, we can write the relation between $\eta_{\Delta L}$ and $\eta_{\Delta B}$ as:
	\begin{eqnarray}
		\begin{aligned}
		\eta_{\Delta L}&=-\eta_{\Delta_{B-L}} \\
		\eta_{\Delta B}(\infty)&=\frac{28}{79}\eta_{\Delta_{ B-L}}(\infty)=-\frac{28}{79}\eta_{\Delta L}(\infty)
	    \end{aligned} 
		\label{eqnarr:asym}
	\end{eqnarray} 
	We now show the obtained analytical value of $\eta_{\Delta B}(\infty)$ (or equivalently $\eta_{\Delta B}^\infty$) for NO and IO cases respectively in figures \ref{fig:11} and~\ref{fig:11I} setting to the benchmark values for all the relevant parameters. For the scatter plot in Fig.~\ref{fig:11}, we obtain $\epsilon_1$ values in a $m_1-\delta_{\text{CP}}$ plane with $m_1$ ranging from $10^{-4}\text{ eV}$ to $1\text{ eV}$ on the X-axis and $\delta_{\text{CP}}$ ranging from $0$ to $2\pi$ on the Y-axis. In the plot, dotted black mesh region implies the parameter space where we obtain a negative sign in the final baryon abundance value ($\eta_{\Delta B}^{\infty}$). It can be seen that within the currently accepted $1\sigma$ value of $\delta$, we obtain a fairly close numerical value with a positive sign of final baryon asymmetry, consistent with the observed value of $\eta_{\Delta B}^{\infty}=(6.105^{+0.086}_{-0.081})\times10^{-10}$~\cite{Planck:2015fie} for our benchmark value of $m_{N_1}=10^{13}\text{ GeV}$, $m_{N_2}=10^{14}\text{ GeV}$, $m_{N_3}=5\times10^{14}\text{ GeV}$ and $m_1=0.01\text{ eV}$. Now, for the scatter plot in Fig.~\ref{fig:11I}, we obtain $\epsilon_1$ values in a $m_3-\delta_{\text{CP}}$ plane with $m_3$ ranging from $10^{-4}\text{ eV}$ to $1\text{ eV}$ on the X-axis and $\delta_{\text{CP}}$ ranging from $0$ to $2\pi$ on the Y-axis. The dotted black mesh region here too implies a negative sign in the obtained~$\eta_{\Delta B}^{\infty}$ value. From the plot, one may see that for the benchmark $m_3$ value of $10^{-2}\text{ eV}$, the value of $\eta_{\Delta B}^{\infty}$ within the $1\sigma$ band of currently accepted $\delta_{\text{CP}}$ value for the inverted ordering lies in the dotted red band region and thus the value of final baryon asymmetry for the benchmark analysis in the IO case is obtained with a negative sign.
	
	Now we present the detailed Boltzmann analysis of the asymmetry evolution for the two cases of NO and IO active neutrino mass hierarchy separately:
	\subsubsection{Normal Ordering}
	\label{subsec:NO2}
	The numerical solution to BEs for the NO case is plotted in Fig.~\ref{fig:10}. Here, right-handed neutrino mass hierarchy used is: $m_{N_1}=10^{13}~\text{GeV}$, $m_{N_2}=10^{14}~\text{GeV}$ and $m_{N_3}=5\times10^{14}~\text{GeV}$. For such a hierarchy, we can safely neglect the contributions of $N_{2,3}$ decay to asymmetry, as any asymmetry created from their decay is washed out by the time when the Universe reaches $T=m_{N_1}$. For our analysis, we assume zero initial asymmetry for leptons, and also we begin with a negligible $N_1$ number density $(N_1(z\rightarrow0)=0)$ instead of a thermal abundance $(N_1(z\rightarrow0)=N_1^{eq}(z\rightarrow0))$. However, we see that our initial conditions do not necessarily change the final asymmetry value, and the same results are obtained for a thermal $N_1$ abundance, too. From the plots, one may see that the overall dynamics of all the interactions play out in such a way that a remnant asymmetry, as required, is present in the leptonic sector. We then assume that this asymmetry is successfully transferred to the baryons via sphalerons, and the obtained final baryon asymmetry is close to the required value.
	
		\begin{figure}
		\centering
		
		\begin{subfigure}{0.47\textwidth}
			\includegraphics[width=\linewidth]{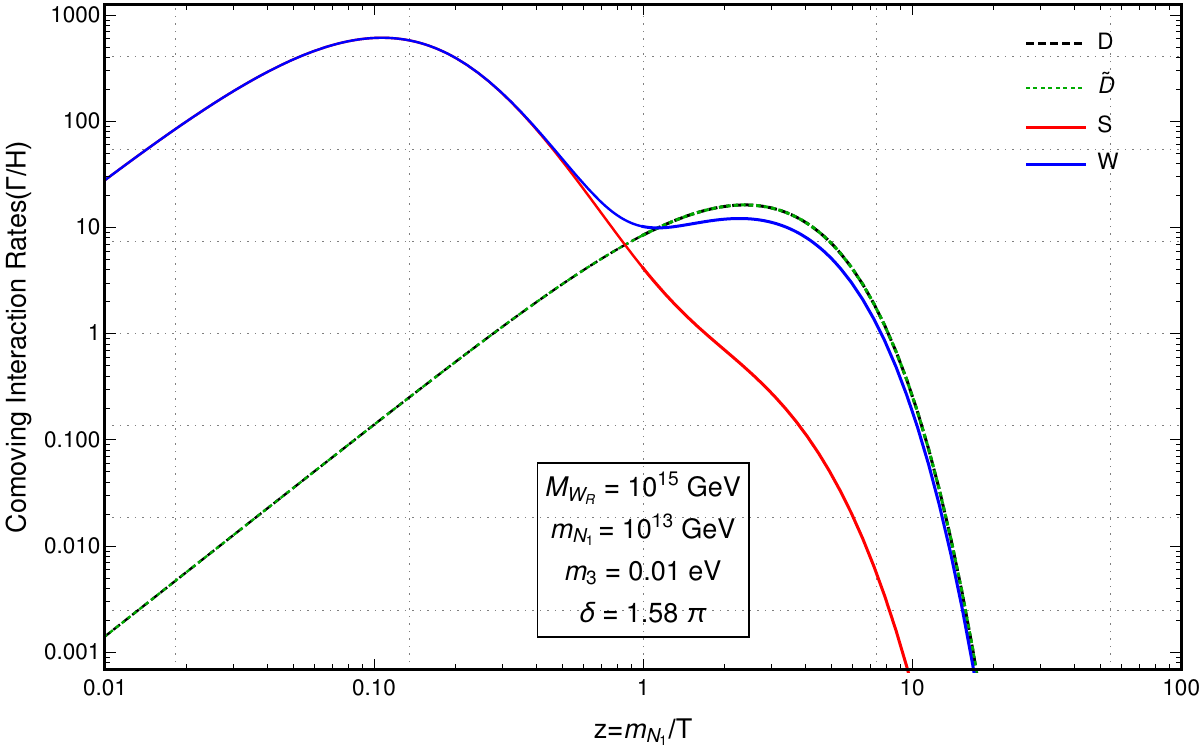}
			\caption{}
			\label{fig:10Isub1}
		\end{subfigure}
		\hfill
		\begin{subfigure}{0.48\textwidth}
			\includegraphics[width=\linewidth]{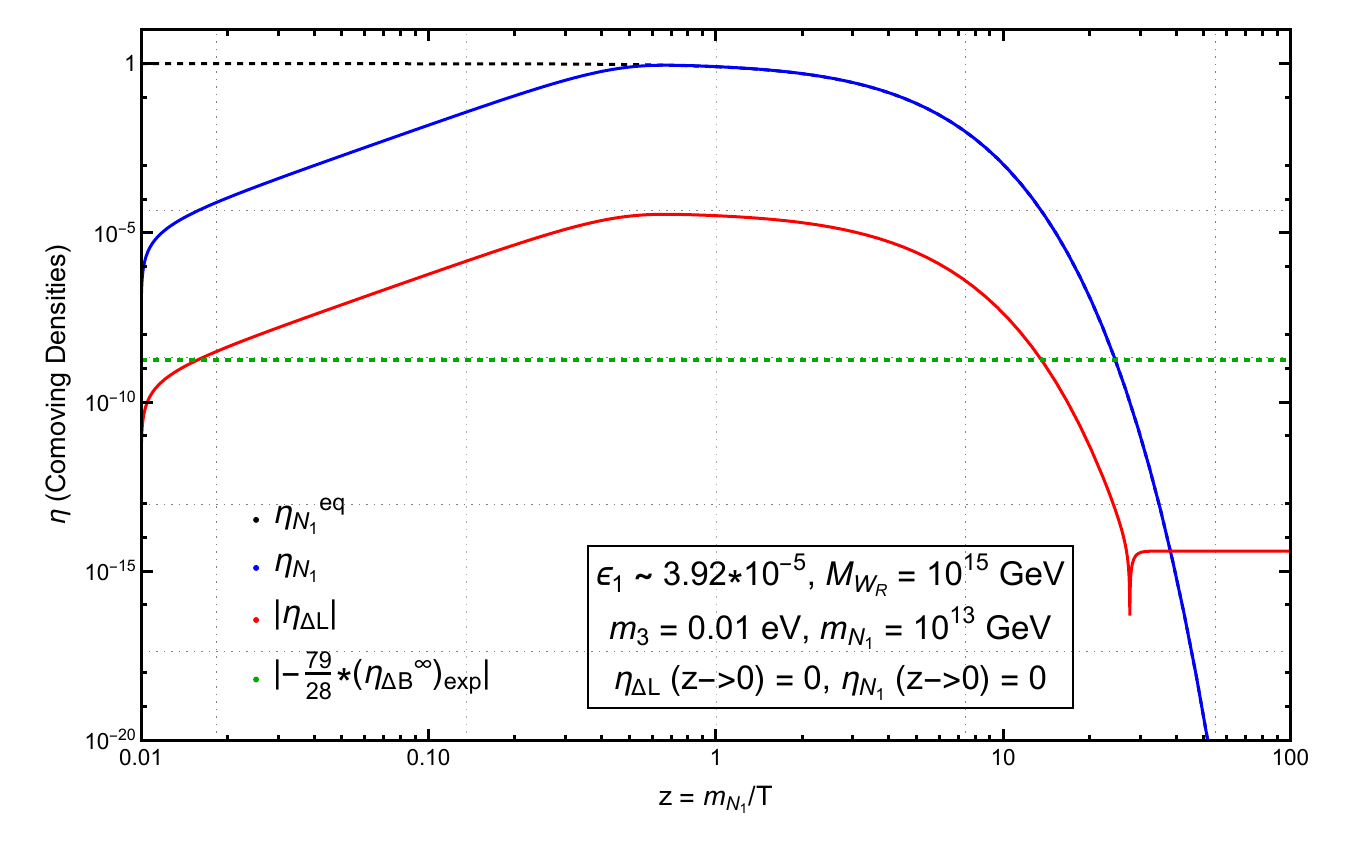}
			\caption{}
			\label{fig:10Isub2}
		\end{subfigure}
		
		\medskip
		
		\begin{subfigure}{0.48\textwidth}
			\includegraphics[width=\linewidth]{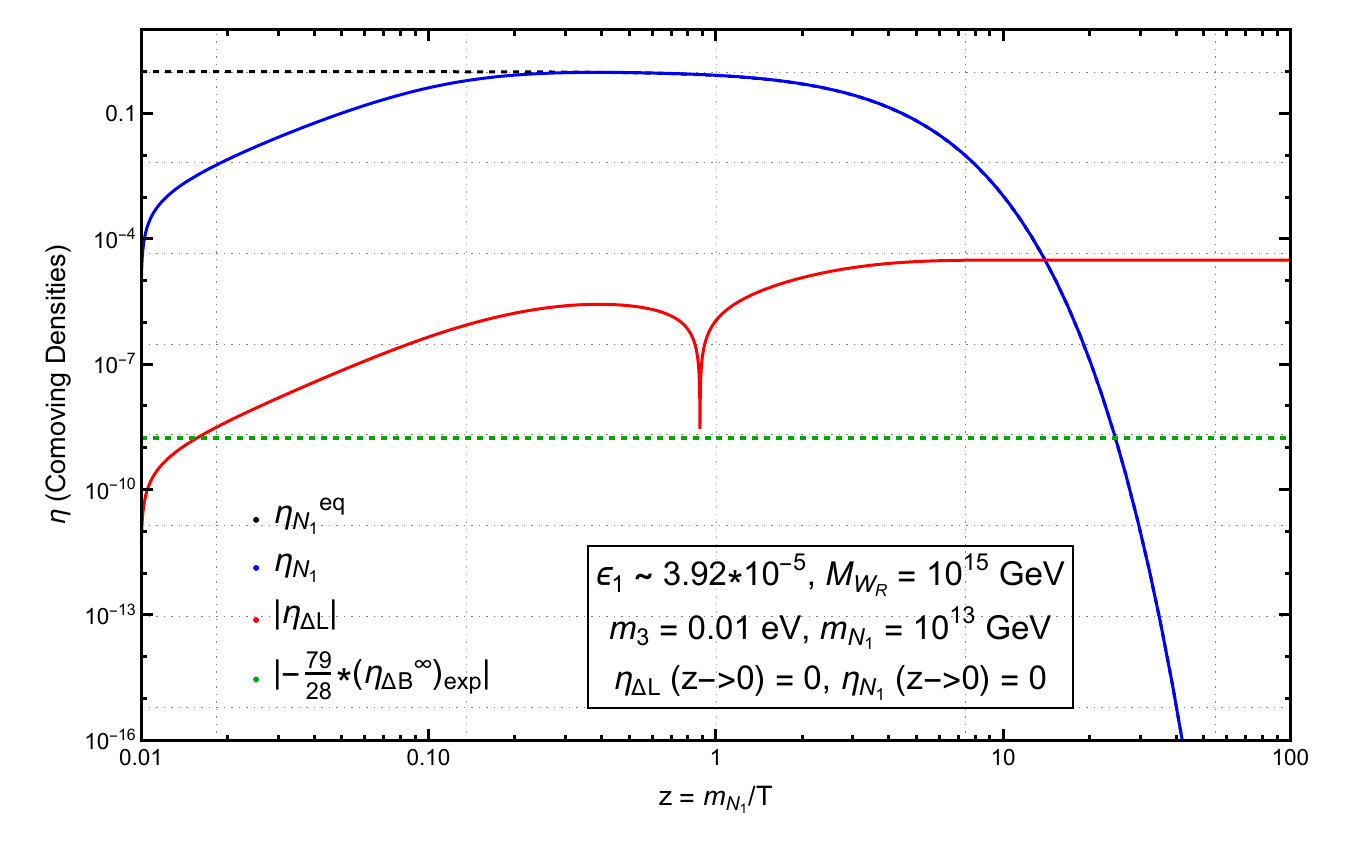}
			\caption{}
			\label{fig:10Isub3}
		\end{subfigure}
		\hfill
		\begin{subfigure}{0.48\textwidth}
			\includegraphics[width=\linewidth]{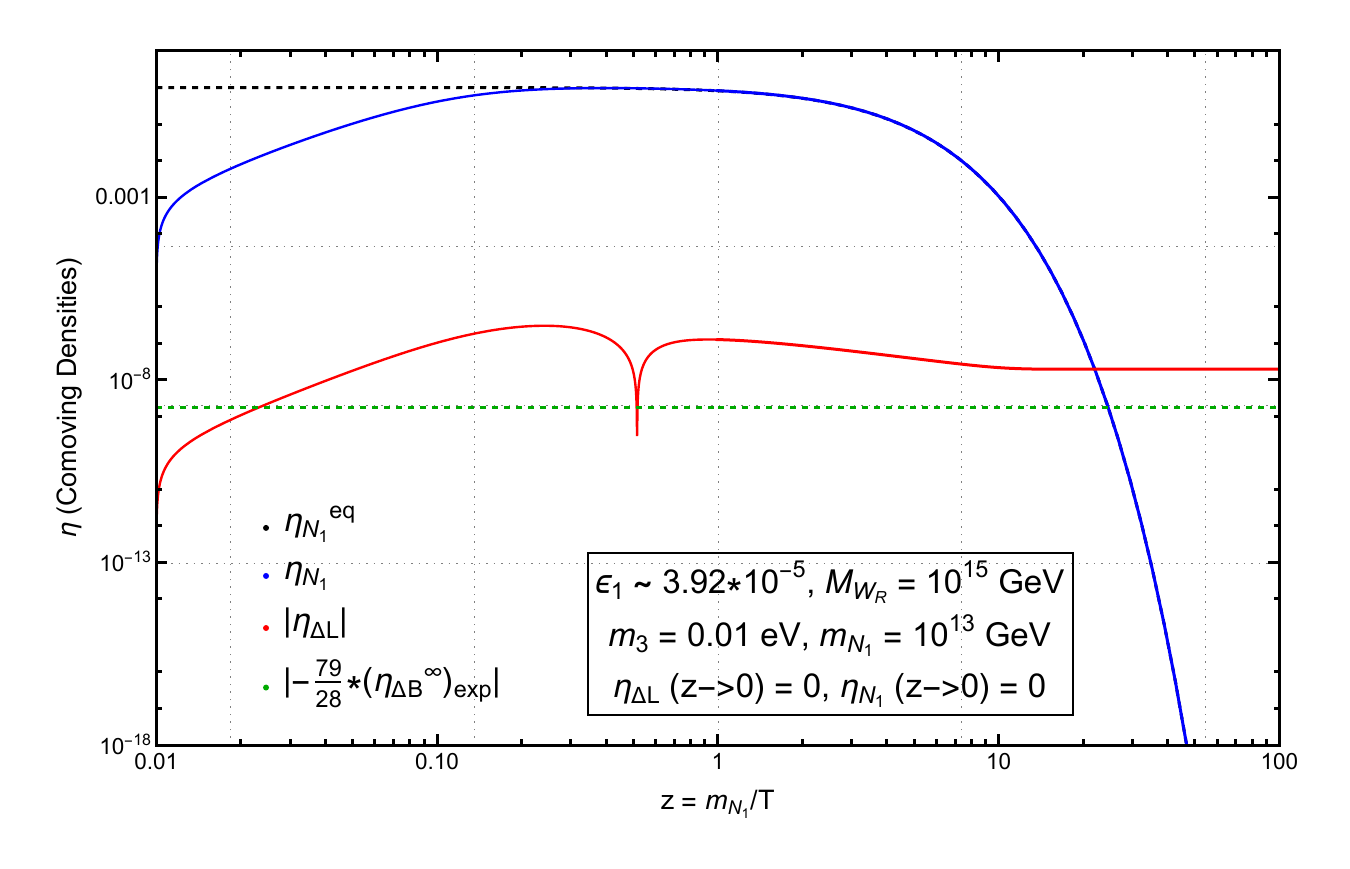}
			\caption{}
			\label{fig:10Isub4}
		\end{subfigure}
		
		\caption{Figure shows four plots depicting the cosmological evolution of the various interaction terms and the number densities present in the coupled BEs (\ref{be1}) and (\ref{be2}) for the \textbf{IO} case. Plot (a) here shows the relative contributions of the different interaction terms that play a role in the evolution of lepton asymmetry. These terms include the $N_1$ decays~(black dashed line, denoted by D), inverse-decays~(green dotted line, denoted by $\tilde{D}$), scatterings involving leptons~(solid red line, denoted by S) and the asymmetry washout interactions~(solid blue line, denoted by W). Plot (b) shows the cosmological evolution of the lepton asymmetry number density with only decays and inverse-decays of $N_1$ considered. In Plot (c), we also include the effects of various scattering interactions involving leptons, while calculating the evolution of lepton asymmetry. Plot (d) shows the cosmological evolution of lepton asymmetry incorporating all the interactions present in the BEs i.e. decays, inverse-decays, scatterings and the terms that lead to asymmetry wash-out. The values of all the relevant input parameters and the considered initial conditions are mentioned within the plots. In the plots (b), (c) and (d), for a reference, we also show the cosmological evolution of the $N_1$ number density~($\eta_{N_1}$) represented by the solid blue line, the $N_1$ equilibrium number density~($\eta_{N_1}^{\text{eq}}$) represented by the dashed black line. In these 3 plots, the magnitude of the number density for the lepton asymmetry value~($|\eta_{\Delta\text{L}}|$) is denoted by the solid red line and the green dashed line represent the currently accepted asymmetry value by observations~\cite{Planck:2015fie}. All the oscillation parameters are fixed at their best-fit values.}
		\label{fig:10I}
	\end{figure}
	
	Analysing the results from Fig.~\ref{fig:10}, firstly we see in the Fig.~\ref{fig:10sub1}, the relative strengths of various interactions that take part in our BEs. The washout interactions shown in the blue line can be seen to lie in strong regime for initial $z$ values $(\Gamma_W(z\rightarrow0)\simeq55))$. In the remaining plots i.e. in figures \ref{fig:10sub2},~\ref{fig:10sub3} and~\ref{fig:10sub4}, we compare numerically the effect of each of those terms in deciding the final asymmetry. The values of relevant input and derived parameters for our analysis here are given as: $m_{N_1}=10^{13}~\text{GeV}$, $m_{N_2}=10^{14}~\text{GeV}$, $m_{N_3}=5\times10^{14}~\text{GeV}$, $m_{W_R}=1\times10^{15}~\text{GeV}$, $\delta=1.08\pi$, $m_1=0.01~\text{ eV}$. The derived value of asymmetry parameter used here is $\epsilon_1\sim-3.8\times10^{-4}$ and the initial conditions that we assume are $\eta_{\Delta L}(z\rightarrow0)=0$ and $\eta_{N_1}(z\rightarrow0)=0$. The decays of $N_1$ alone are quite insufficient, as seen in Fig.~\ref{fig:10sub2}, where the evolution of asymmetry in the lepton sector follows the rate of $N_1$ decay. At a very late stage$(z\sim28)$, the asymmetry value changes sign. This could be explained by the relative strength of $N_1$ number density with respect to $N_1^{eq}$ number density. $N_1$ number density slightly increases at around $z\sim27$. This changes the sign of $\left(\frac{\eta^{N_1}}{\eta^N_{\rm eq}}-1\right)$ term in BEs, the effect of which is seen in the asymmetry evolution equation. The final lepton asymmetry value remains quite underabundant here $(\eta_{\Delta L}^{\infty}\sim5\times10^{-17})$. The inclusion of scattering interactions with these decays, as shown in Fig.~\ref{fig:10sub3} provides a final asymmetry that is more than the required value by around $10^4$ times. Here, the change of sign for asymmetry occurs at $z\sim1.30$, and after that, due to the lepton number violating scatterings of $N_1$, the lepton asymmetry increases quite significantly, as we do not include washout interactions here to suppress their effects. In the final plot i.e. Fig.~\ref{fig:10sub4}, we see the effects of all the interactions, including the washout terms. The overall interplay and dynamics of these terms provide an acceptable final asymmetry in the lepton sector close to $(\eta_{\Delta L}^{\infty}\sim-6\times10^{-10})$. Then, we assume a successful transfer of this asymmetry to the baryonic sector via sphalerons. As the value of asymmetry parameter,~($\epsilon_1$) comes to be negative in this NO case~(Eq.~\ref{eq:asymNO}), the sign of final baryon asymmetry,~($\eta_{\Delta B}^{\infty}$) is obtained to be positive here, in accordance with the requirement.
	
	\subsubsection{Inverted Ordering}
	\label{subsec:IO2}
	The numerical solution to BEs for the IO case is plotted in Fig.~\ref{fig:10I}. Here, the right-handed neutrino mass hierarchy and the initial conditions are considered to be same as in the NO case given in subsection \ref{subsec:NO2}. From the figure, we see that the overall dynamics of the interactions involved here give us a non-zero asymmetry in the leptonic sector which is assumed to be transferred to the baryonic sector via sphalerons.
	
	Analysing the results from Fig.~\ref{fig:10I}, in the first plot i.e. Fig~\ref{fig:10Isub1} we show the relative strengths of various interactions that take part in the BEs. Here also, we see that the initial washout interaction~(blue line) is in the strong washout regime and thus the washout interactions play an important role in the overall asymmetry evolution dynamics. In the remaining plots i.e. figures \ref{fig:10Isub2},~\ref{fig:10Isub3} and~\ref{fig:10Isub4}, we compare numerically the effect of each of the decay, inverse-decay, scattering and washout terms in deciding the final asymmetry. The values of relevant input and derived parameters for our analysis here are given as: $m_{N_1}=10^{13}~\text{GeV}$, $m_{N_2}=10^{14}~\text{GeV}$ , $m_{N_3}=5\times10^{14}~\text{GeV}$, $m_{W_R}=1\times10^{15}~\text{GeV}$, $\delta=1.58\pi$, $m_3=0.01~\text{ eV}$. The derived value of asymmetry parameter used here is $\epsilon_1\sim3.92\times10^{-5}$ and the initial conditions that we assume are $\eta_{\Delta L}(z\rightarrow0)=0$ and $\eta_{N_1}(z\rightarrow0)=0$. Upon looking at the plot in Fig.~\ref{fig:10Isub2}, we see that the decay interactions alone gives an under-abundant value for final asymmetry in the leptonic sector. Also, we see that at a very late time$(z\sim28)$, the value of leptonic asymmetry~(red line) changes its sign. This could be explained by the relative strength of $N_1$ number density with respect to $N_1^{eq}$ number density. $N_1$ number density slightly increases at around $z\sim27$. This changes the sign of $\left(\frac{\eta^{N_1}}{\eta^N_{\rm eq}}-1\right)$ term in BEs, the effect of which is seen in the asymmetry evolution equation. The overall dynamics for the rest of the plots (figures \ref{fig:10Isub3} and \ref{fig:10Isub4}) remains same as that for the NO case and thus we avoid putting it here again, so as to avoid repetition. One may refer subsection~\ref{subsec:NO2} for the detailed explanation of these plots behaviour. Now, as the value of asymmetry parameter,~($\epsilon_1$) comes to be positive in this IO case~(Eq.~\ref{eq:asymIO}), the sign of final baryon asymmetry,~($\eta_{\Delta B}^{\infty}$) for the given parameter space is obtained to be negative here. Thus one may repeat the similar analysis for a different parameter space in the IO case, where the $\epsilon_1$ value is obtained to be negative, so as to ensure a correct magnitude as well as sign value for the final baryon asymmetry.
	
	After numerically analysing the asymmetry evolution for both the cases under study  i.e. NO and IO, we see that in the NO case~(within the considered parameter space), the CP-violating phase $(\delta)$ alone (as we consider zero Majorana phases for the Boltzmann analysis) is sufficient to provide the required baryon asymmetry ($\eta_{\Delta B}^{\infty}$) with its correct sign confirming the main theme in our work of connecting the low-energy CP violation and matter-antimatter asymmetry of the Universe. In the IO case, we see that for a similar parameter space the value for the obtained~$\eta_{\Delta B}^{\infty}$ gets a negative sign and thus we need to choose a different parameter space in this case to satisfy the successful baryogenesis requirements. One may note that in our model framework, a Boltzmann analysis with non-zero Majorana phases would not alter the overall results, as the dependence of the CP asymmetry parameter ($\epsilon_1$) on Majorana phases is negligible for benchmark point input values. This dependence has been explored in subsections \ref{subsec:NO1} and \ref{subsec:IO1}. Nevertheless, using this setup, a future confirmation of low-energy CP-violating phases will lead to a direct estimation of BAU. We are thus keen on the results obtained in the next 10-15 years from some of the ongoing long-baseline oscillation experiments like NOvA~\cite{NOvA:2019cyt}, T2K collaboration~\cite{T2K:2019bcf}, DUNE~\cite{Esteban:2020cvm} and also future projects like JUNO, which will provide us a clearer picture of this CP phase, making our work relevant and important.
	
	\section{Conclusion}
	\label{sec:conc}
	
	We have investigated the scenario of thermal unflavored leptogenesis within a class of Left-Right Symmetric Models with a scalar bidoublet and doublets while adding a single copy of sterile neutrino, $S_L$ per generation in the fermion sector. In the absence of singlet fermion, the light neutrinos are Dirac in nature with no Majorana mass term for right-handed neutrinos $(N_R)$, thus prohibiting leptogenesis. The interesting feature of double seesaw with a large Majorana mass term for $S_L$ generates Majorana masses for the left-handed $(\nu_{L})$ as well as right-handed neutrinos $(N_{R})$. Thus, the resulting lepton number violating out-of-equilibrium decays of $N_R$ can successfully generate the required CP asymmetry in the lepton sector, which can be converted to the baryon asymmetry of the Universe.
	
	The source of CP violation required for successful leptogenesis is the generic Dirac neutrino mass matrix, $(M_D)$, which connects $\nu_{L}$ and $N_R$. The structure of $M_D$ plays a vital role as the CP asymmetry parameter, $\epsilon_1$ is determined from the imaginary part of $M_D^{\dagger}M_D$. The considered screening condition, along with the discrete LR symmetry $C$ assists us to derive the structure of $M_D$ in terms of light and heavy RH neutrinos and light neutrino mixing matrix $U_{PMNS}$. Thus, the uniqueness of our work lies in the fact that $M_D$ becomes dependent on low-energy CP-violating Dirac phase, $\delta$ (contained in $U_{PMNS}$) without invoking any parametrization. This provides an exciting motivation for current long baseline experiments like NOvA, T2K, DUNE, T2HK, T2HKK and also future projects like JUNO to probe leptogenesis indirectly through $\delta$.
	
	By taking the best-fit values for $\delta$ and other low-energy oscillation parameters of $U_{PMNS}$, along with the mass of light neutrino within current experimental bounds and the eigenvalues of RH neutrino mass matrix as input parameters, we completely determine the matrix structure of $M_D$. For comprehensiveness, we perform our analysis for both the normal and inverted mass orderings of the light neutrino. The mass range of right-handed neutrinos~$(N_i)$ required to perform a thermal leptogenesis comes to be $m_{N_1}\geq10^{13}~\text{GeV}$. Such a choice serves a two-fold purpose here. It ensures we can safely neglect flavor effects in the heavy neutrino sector and allows for a significant asymmetric decay of these $N_1$ to generate an effective asymmetry, $\epsilon_1$. With the considered benchmark points for the NO case, we obtain a value of final baryon asymmetry consistent with the observed value, $\eta_{\Delta B}^{\infty}=(6.105^{+0.086}_{-0.081})\times10^{-10}$. For the similar benchmark values, the final asymmetry obtained in the IO case has roughly the same order of magnitude but with a minus sign.
	
	In our analysis, we find an interesting outcome that within the considered model framework of double seesaw, the value of asymmetry parameter $(\epsilon_1)$ exhibits negligible dependence on the Majorana phases $\alpha$ and $\beta$ for the given set of input parameters in both the NO and IO cases. This highlights $\delta$ as the prime source for generating the required baryon asymmetry. Nevertheless, for some other choice of input parameters, one may obtain a distinct dependence of $\epsilon_1$ on $\alpha$ and $\beta$ but such a choice might deviate us from the thermal unflavoured regime. So, we plan on extending this work to study the impact of non-zero Majorana phases in the flavoured or resonant regime of leptogenesis, thus allowing the testability of the framework via neutrinoless double beta decay in near-future experiments. Other leptogenesis scenarios, like Dirac, could also be considered within this setup to bring down the scale of leptogenesis itself. This, in turn, can have new implications for the mass bounds on $N_R$ and right-handed gauge boson, $W_R$.

	\section*{Acknowledgement}
	Utkarsh Patel~(UP) and Pratik Adarsh~(PA) would like to acknowledge the financial support obtained from the Ministry of Education, Government of India. Purushottam Sahu~(PS) would like to acknowledge the Institute Postdoctoral Fellowship of IIT Bombay for financial support. PS also acknowledges the support from the Abdus Salam International Centre for Theoretical Physics~(ICTP) under the "ICTP Sandwich Training Educational Programme~(STEP)" SMR.3676 and SMR.3799, where part of the analysis was performed during his stay. We also thank Dr.~Alessandro Granelli for his valuable insights and comments on the implications of our framework.
	
	\newpage
	\appendix	
	\section{Order of Magnitude for Mass Parameters }\label{ordermagApp}
	We present the order of magnitude for various mass parameters by generalizing Table \ref{ordermag} in the broader context of leptogenesis.
	\setlength{\bigstrutjot}{6pt}
	\begin{table}[!h]
		\centering
		\begin{tabular}{|c|c||c|c|c||c|c|c|}
			\hline
			\textbf{Regime} & \textbf{Flavor} & $\mathbf{M_D}$ & $\mathbf{M_{RS}}$ & $\mathbf{M_S}$ & $\mathbf{m_\nu}$\textbf{(eV)} &$\mathbf{m_N}$ &$\mathbf{m_S}$ \\
			\hline
			\hline
			\multirow[c]{6}{*}[-30pt]{\rotatebox[origin=c]{90}{\textbf{Thermal}}} & \multirow[c]{3}{*}[-6pt]{\rotatebox[origin=c]{90}{\textit{\textbf{Unflavored}}}}\bigstrut[t]\bigstrut[b] &	$10$  & $10^{14}$  & $10^{15}$ & $0.01$  & $10^{13}$ & $10^{15}$ \\
			\cline{3-8}
			& 1\bigstrut[t]\bigstrut[b] & $10^2$  & $10^{15}$  & $10^{16}$ & $0.1$  & $10^{14}$ & $10^{16}$ \\
			\cline{3-8}
			& & $1$ & $10^{13}$ & $10^{14}$\bigstrut[t]\bigstrut[b] & $0.001$ & $10^{12}$ & $10^{14}$\\
			\cline{2-8}
			& \multirow[c]{3}{*}[-10pt]{\rotatebox[origin=c]{90}{\textit{\textbf{Flavored}}}}\bigstrut[t]\bigstrut[b] & $3.2\times10^{-3}$ & $10^7$ & $10^8$ & $0.01$ & $10^6$ & $10^8$ \\
			\cline{3-8}
			& & $3.2\times10^{-2}$ & $10^8$ & $10^9$\bigstrut[t]\bigstrut[b] & $0.1$ & $10^7$ & $10^9$ \\
			\cline{3-8}
			& & $10^{-1}$ & $10^{9}$ & $10^{10}$ \bigstrut[t]\bigstrut[b] & $0.1$ & $10^8$ & $10^{10}$\\
			\hline
			\multirow[c]{3}{*}[-9pt]{\rotatebox[origin=c]{90}{\textbf{Resonant}}} & \multirow[c]{3}{*}[-6pt]{\rotatebox[origin=c]{90}{\textit{\textbf{Unflavored}}}}\bigstrut[t]\bigstrut[b] & $10^{-4}$  & $10^{3}$  & $10^{4}$ & $0.1$  & $10^{2}$ & $10^{4}$ \\
			\cline{3-8}
			& \bigstrut[t]\bigstrut[b] & $10^{-5}$  & $10^{2}$  & $10^{3}$ & $0.01$  & $10$ & $10^{3}$ \\
			\cline{3-8}
			& \bigstrut[t]\bigstrut[b] & $10^{-5}$  & $10^{1}$  & $10^{2}$ & $0.1$  & $1$ & $10^{2}$ \\
			\hline
		\end{tabular}
		\caption{Order of magnitude estimation of various neutrino masses in LRSM with double seesaw mechanism. The values are estimated for both resonant and thermal regime with flavored and unflavored cases. All masses except the active neutrino masses are in GeV.}
		\label{ordermagAppTab}
	\end{table}
	\section{Derivation of $M_D$}\label{AppendixMD}
	To derive the final expression of $M_D$ as in Eq. (\ref{MDexpression}), we start with Eq. (\ref{MRSfirst}) and (\ref{screeningNew}) such that
	\begin{equation}\label{MDappendix}
		M_D=i.m_\nu\sqrt{m_\nu^{-1}m_N}.
	\end{equation}
	We note that the light neutrino and heavy neutrino mass matrices are diagonalized with same mixing matrix $U_\nu$ as per relation obtained in the Eq. (\ref{UNUnu}). Therefore, we have 
	\begin{eqnarray}\label{mnumnuinvmN}
		\begin{aligned}
			m_\nu&=U_\nu \hat{m}_\nu U_\nu^T\,;\\
			m_\nu^{-1}&=U_\nu^* 	\hat{m}_\nu^{-1}U_\nu^\dagger\,;\\
			m_N&=U_\nu \hat{m}_N U_\nu^T\,.
		\end{aligned}
	\end{eqnarray}
	Putting relations from Eq. (\ref{mnumnuinvmN}) into the Eq. (\ref{MDappendix}) and noting that $U_\nu^\dagger U_\nu =\mathds{1}$ as $U_\nu$ is unitary, we have
	\begin{eqnarray}\label{MDappendix2}
		\begin{aligned}
			M_D&=i.U_\nu \hat{m}_\nu U_\nu^T\sqrt{U_\nu^* 	\hat{m}_\nu^{-1}\hat{m}_N U_\nu^T}\\
			&=i.U_\nu \hat{m}_\nu U_\nu^T\sqrt{U_\nu^* 	D U_\nu^T}\,.
		\end{aligned}
	\end{eqnarray} Here $D(=\hat{m}_\nu^{-1}\hat{m}_N)$ is a diagonal matrix. A non-diagonal symmetric matrix (say $A$) can always be diagonalized by using a similarity transformation as:
	\begin{equation*}
		A=PBP^{-1}.
	\end{equation*}
	Here $P$ is a non-singular matrix, and $B$ is a diagonal matrix. We also have this lemma that for any $m\in Q$,
	\begin{equation}\label{lemma}
		A^m=PB^mP^{-1}.
	\end{equation}
	Using Eq. (\ref{lemma}) in Eq. (\ref{MDappendix2}) and noting that the non-singular matrix $P$ for our case is also unitary ($U_\nu$), we have
	\begin{equation}\label{MDappendix3}
		M_D=i.U_\nu \hat{m}_\nu\underbrace{U_\nu^T U_\nu^*}_{\mathds{1}} 	D^{1/2} U_\nu^T\,.
	\end{equation}
	We also note that $D(=\hat{m}_\nu^{-1}\hat{m}_N)$ is real. Hence $D^*=D$ and therefore finally our $M_D$ becomes:
	\begin{equation}\label{MDappendixFinal}
		M_D=	i.U_\nu\hat{m}_\nu(\hat{m}_\nu^{-1}\hat{m}_N)^{1/2}U_\nu^T\,.
	\end{equation}
	\section{Interaction Rates}
	\label{app:ir}
	
	\subsection{Compact Form}
	The various decay and scattering rates involved in BEs~(\ref{be1}) and (\ref{be2}) are given below in terms of the physical decay and scattering parameters involving the heavy neutrinos:
	\begin{align}
		\gamma^D_{l\alpha} \ & = \ \gamma^{N_\alpha}_{L_l\phi_l} + \underbrace{\gamma^{N_\alpha}_{l_Rq\bar{q}'}}_{\text{Extra effective 3-body decay in LRSM due to }W_R},
		\label{decay} \\
		\tilde{\gamma}^D_{l\alpha} \ & = \ \gamma^{N_\alpha}_{L_l\phi_l}, \label{decay2}\\
		\gamma^{S_L}_{l\alpha} \ & = \ \underbrace{\gamma^{N_\alpha L_l}_{Qu^c}+\gamma^{N_\alpha u^c}_{L_l Q^c} + \gamma^{N_\alpha Q}_{L_l u}}_{{\text{Scalar Mediated}}} + \underbrace{\gamma^{N_\alpha L_l}_{\phi^\dag V_\mu} + \gamma^{N_\alpha V_\mu}_{L_l \phi} + \gamma^{N_\alpha \phi^\dag}_{L_l V_\mu}}_{{\text{Gauge Bosons Mediated}}}, \label{scatsl1}\\
		\tilde{\gamma}^{S_L}_{l\alpha} \ & = \ \frac{\eta^N_\alpha}{\eta^N_{\rm eq}}\gamma^{N_\alpha L_l}_{Qu^c}+\gamma^{N_\alpha u^c}_{L_l Q^c} + \gamma^{N_\alpha Q}_{L_l u} + \frac{\eta^N_\alpha}{\eta^N_{\rm eq}}\gamma^{N_\alpha L_l}_{\phi^\dag V_\mu} + \gamma^{N_\alpha V_\mu}_{L_l \phi} + \gamma^{N_\alpha \phi^\dag}_{L_l V_\mu},\label{scatsl2} \\
		\gamma^{S_R}_{l\alpha} \ & = \ 
		\underbrace{\gamma^{N_\alpha l_R}_{\bar{u}_R d_R} + \gamma^{N_\alpha \bar{u}_R}_{l_R \bar{d}_R} + \gamma^{N_\alpha d_R}_{l_R u_R}}_{{\text{Extra term in LRSM: }}{W_R\text{ Gauge Boson Mediated}}},\label{scatsr1}\\
		\tilde{\gamma}^{S_R}_{l\alpha} \ & = \ 
		\frac{\eta^N_\alpha}{\eta^N_{\rm eq}} \gamma^{N_\alpha l_R}_{\bar{u}_R d_R} + \gamma^{N_\alpha \bar{u}_R}_{l_R \bar{d}_R} + \gamma^{N_\alpha d_R}_{l_R u_R},\label{scatsr2}\\
		\gamma^{(\Delta L=2)}_{lk} \ & = \ \underbrace{\gamma'^{L_l\phi_l}_{L_k^c\phi_k^\dag}+\gamma^{L_l L_k}_{\phi_l^\dag \phi_k^\dag}}_{{\text{terms leading to depletion of lepton number density}}},\label{scatdl2} \\
		\gamma^{(\Delta L=0)}_{lk} \ & = \ \gamma'^{L_l\phi_l}_{L_k\phi_k}+\gamma^{L_l\phi_l^\dag}_{L_k\phi_k^\dag}+\gamma^{L_l L_k^c}_{\phi_l \phi_k^\dag} \label{scatdl0}\; .
	\end{align}
	In the above equations, all the $\gamma$ terms with $\sim$ and $\gamma^{(\Delta L=0,2)}_{lk}$ terms i.e.~\cref{decay2,scatsl2,scatsr2,scatdl0,scatdl2} are corresponding to the {collision term} in second Boltzmann equation i.e. (\ref{be2}). The scattering terms involving two heavy neutrinos in the initial state, e.g. induced by a $t$-channel $W_R$ or $e_R$, and by an $s$-channel $Z_R$, are not included here since their rates are doubly Boltzmann-suppressed and numerically much smaller than the scattering rates given above~\cite{Frere:2008ct,Blanchet:2009bu,Blanchet:2010kw}.

	\subsection{Explicit Form}
	The decay rates in equations \ref{decay} and \ref{decay2} are explicitly given by:
	\begin{eqnarray}
		\begin{aligned}
			\gamma^{N_1}_{L_l\phi} \ & = \ \frac{m_{N_1}^3}{\pi^2 z}K_1(z)\left[\Gamma(N_1\to L_l \phi)+\Gamma(N_1\to L_l^c\phi^\dag)\right], \label{drates1}\\
			\gamma^{N_1}_{l_Rq\bar{q}'} \ & = \ \frac{m_{N_1}^3}{\pi^2 z} K_1(z)\left[\Gamma(N_1\to l_R q_R\bar{q}'_R)+\Gamma(N_1\to \bar{l}_R\bar{q}_R q'_R)\right] \label{drates2}\; .
		\end{aligned}
	\end{eqnarray}
	In the above equations, the explicit expressions for various $\Gamma$'s are given as:
	\begin{eqnarray}
		\begin{aligned}
			\Gamma(N_{1}\to L_l\phi) \  &= \  \frac{m_{N_1}}{16\pi}Y_{D_{1 1}}{Y^*_{D_{1 1}}}\; , \qquad   \Gamma(N_{1}\to L^c_l\phi^\dagger ) \
			= \  \frac{m_{N_1}}{16\pi}{Y^c_{D_{1 1}}}{Y^{c*}_{D_{1 1}}}\; .
			\label{gamma}
		\end{aligned}
	\end{eqnarray}
	\begin{multline}
		\Gamma(N_1\to l_R q_R\bar{q}'_R) \ = \ \Gamma(N_1\to \bar{l}_R \bar{q}_R q'_R) \ = \ \frac{3g_R^4}{2^9\pi^3 m_{N_1}^3}\nonumber\\ \times \int_0^{m_{N_1}^2} ds \frac{m_{N_1}^6-3m_{N_1}^2 s^2+2 s^3} {(s-M_{W_R}^2)^2+M_{W_R}^2\Gamma_{W_R}^2} \; ,
		\label{3body}
	\end{multline}
	where $\Gamma_{W_R}\simeq (g_R^2/4\pi)M_{W_R}$ is the total decay width of $W_R$.
	
	Other than the decay processes, there are scatterings involving $N_1$ and other leptons which play a role in asymmetry evolution. The general expression for $\gamma$ of a $2\leftrightarrow 2$ scattering process as given in reference~\cite{BhupalDev:2014hro} , $XY\leftrightarrow AB$ can be defined as:
	\begin{eqnarray}
		\gamma^{XY}_{AB} \ = \ \frac{m_{N_1}^4}{64\pi^4 z}\int _{x_{\rm thr}}^\infty dx\sqrt{x} K_1(z\sqrt{x})\hat{\sigma}^{XY}_{AB}(x), \label{scat}
	\end{eqnarray}
	where $x=s/m_{N_1}^2$ with the kinematic threshold value $x_{\rm thr} = {\rm max}[(m_X+m_Y)^2,(m_A+m_B)^2]/m_{N_1}^2$, and $\hat{\sigma}^{XY}_{AB}(x)$ are the relevant reduced cross sections. The scattering rates mentioned in equations (\ref{scatsl1}) to (\ref{scatdl0}) are all to be calculated using this general expression~(\ref{scat}) for a $2\leftrightarrow 2$ scattering process. The explicit reduced cross-section rate expressions for these various scattering processes can be found in references \cite{Pilaftsis:2003gt, Luty:1992un, Frere:2008ct, Blanchet:2010kw, Pilaftsis:2005rv, Chauhan:2021xus, Giudice:2003jh}.
	
	\clearpage
	\bibliographystyle{utcaps_mod}
	\bibliography{Draft_V3}
\end{document}